\documentclass[11pt,notitlepage]{article}

\usepackage{cite,color,amsmath,amssymb,color,fancybox,wrapfig,url,paralist,subfigure,float,verbatim,varioref,amsthm,changepage}
\usepackage{graphicx}

\usepackage[font={small}]{caption}

\usepackage{geometry,setspace,times,ifthen}
\usepackage{quoting}
\quotingsetup{vskip=1pt}

\usepackage{pdfpages}

\usepackage{fancyhdr}

\usepackage[compact]{titlesec}

\usepackage{tabto}

\usepackage{tikz,pgfplots}
\usepackage[european resistor, european voltage, european current]{circuitikz}
\usetikzlibrary{arrows,shapes,positioning,snakes}
\usetikzlibrary{decorations.markings,decorations.pathmorphing,decorations.pathreplacing}
\usetikzlibrary{calc,patterns,shapes.geometric,topaths,fit}

\newtheorem{thm}{Theorem}

\newenvironment{myenv}{\begin{adjustwidth}{0.5cm}{}}{\end{adjustwidth}}

\newcommand{\red}[1] {\textcolor{red}{#1}}
\renewcommand{\thepage}{\arabic{page}}

\fancypagestyle{firststyle}
{
   \fancyhf{}
   \fancyfoot[C]{\footnotesize C-1}
}

\usepackage{mathptmx}

\makeatletter
\def\nl#1#2{\begingroup
    #2%
    \def\@currentlabel{#2}%
    \phantomsection\label{#1}\endgroup
}
\makeatother

{\ensuremath{
\begin{empheq}[box=\fbox]{align}
{#1}
\end{empheq}
}}

\newtheorem{theorem}            {Theorem}[section]

{
\begin{mdframed}
\par\noindent\textbf{#1:}\begin{rmfamily}\noindent}%
{\end{rmfamily}
\end{mdframed}
}

\newcommand{\I}{\Pi}

\newcommand{\pdf}{p}

\newcommand{\prob}{\mathbb{P}}
\newcommand{\E}                 {\Bbb{E}}

\newcommand{\obs}{y}

\newcommand{\state}{x}
\newcommand{\statespace}{\mathcal{X}}
\newcommand{\obspace}{\mathcal{Y}}
\newcommand{\statedim}{X}

\newcommand{\fun}{\phi}

\newcommand{\identity}{I}

\newcommand{\oprob}{B}
\newcommand{\tp}{P}

\newcommand{\btp}{\bar{\tp}}

\newcommand{\finaltime}{N}

\newcommand{\model}{\theta}

\newcommand{\belief}{\pi}

\newcommand{\priv}{\eta}

\newcommand{\gs}{\geq_s}

\newcommand{\filterd}{\sigma}

\newcommand{\filter}{T}

\newcommand{\argmin}{\operatornamewithlimits{argmin}}
\newcommand{\argmax}{\operatornamewithlimits{argmax}}

\newcommand{\reals}{{\rm I\hspace{-.07cm}R}}

\newcommand{\beq}{\begin{equation}}
\newcommand{\eeq}{\end{equation}}
\newcommand{\nn}{\nonumber}

\newcommand{\Y}{\mathbf{Y}}

\renewcommand{\th}{\theta}

\newcommand{\p}{\prime}

\newcommand{\gR}{\succeq}

\newcommand{\bmodel}{\bar{\model}}

\newcommand{\cost}{c}

\newcommand{\reward}{r}

\newcommand{\action}{u}

\newcommand{\discount}{\rho}

 \newcommand{\Ep}{\E_{\policy}}

\newcommand{\policy}{\mu}

\newcommand{\optpolicy}{\policy^*}
\newcommand{\bpolicy}{{\boldsymbol{\mu}}}

\newcommand{\stopset}{\mathcal{S}}

\newcommand{\boprob}{\bar{\oprob}}

\newcommand{\D}{D}

\newcommand{\A}{\mathbb{A}}

\newcommand{\barray}{\begin{array}{ll}}
\newcommand{\earray}{\end{array}}

\newcommand{\e}{\varepsilon}

\newcommand{\degdist}{\rho}

\newcommand{\monprice}{p}

\newcommand{\riskf}{\mathbf{R}}

\newcommand{\pol}{\mu}
\newcommand{\utilitytogo}{J}

\newcommand{\physical}{s}
\newcommand{\laction}{a}

\tikzset{
    blocka/.style={rectangle, draw, line width=0.3mm, black, text width=3.3em, text centered,
                 minimum height=1em},
               line/.style={draw, -latex}}
\tikzset{
    blockd/.style={rectangle, draw, line width=0.2mm, black, text width=10.8em,  text centered,
                 minimum height=2em},
               line/.style={draw, -latex}}
\tikzset{
    blockss/.style={rectangle, draw, line width=0.3mm, black, text width=2.5em, text centered,
                 minimum height=1em},
               line/.style={draw, -latex}}

  \tikzset{
    blockssd/.style={rectangle, draw, line width=0.3mm, black, text width=7.5em, text centered,
                 minimum height=1em},
               line/.style={draw, -latex}}

             \newcommand{\globalpolicy}{\mu}

             \tikzset{
    blockff/.style={rectangle, draw, line width=0.2mm, black,  text centered,text width=7.5em,
                 minimum height=2em},
               line/.style={draw, -latex}}

             \tikzset{
    blockfg/.style={rectangle, draw, line width=0.2mm, black,  text centered,text width=9.5em,
                 minimum height=2em},
               line/.style={draw, -latex}}

\newcommand{\private}{\eta}
\newcommand{\dtime}{n}
\newcommand{\public}{\pi}

\newcommand{\filterg}{\operatorname{Bayes}}
\newcommand{\truevalue}{\bar{f}}

\newcommand{\rednode}{{\tt Red}}
\newcommand{\bluenode}{{\tt Blue}}
\newcommand{\influence}{\mathcal{I}}
\newcommand{\proba}{\operatorname{Prob}}

\newcommand{\filtersocial}{S}

\newcommand{\hg}{\hat{D}}
\newcommand{\tg}{\tilde{D}}
\newcommand{\tim}{n}

\newcommand{\emc}{\rho}
\newcommand{\esa}{\epsilon}
\newcommand{\observ}{Y}
\newcommand{\cvr}{\text{CVaR}_{\alpha}(c(x_{k},a))}

\newcommand{\brho}{\bar{\rho}}

\usepackage[labeled,resetlabels]{multibib}

\newcommand{\citeA}{\cite}
\newcommand{\citeB}{\cite}

\begin{document}

\begin{titlepage}

\title{Dynamics of Social Networks: Multi-agent Information Fusion, Anticipatory Decision Making and  Polling}

\author{ Vikram Krishnamurthy,\\
School of Electrical and Computer Engineering,\\
Cornell University, Ithaca, New York\\
vikramk@cornell.edu
}

\date{}
\maketitle

\begin{abstract}
Social sensors are  agents that provide information about their environment  (state) to a social network  after
interaction  with other social sensors -- this information fusion is  modeled by social learning.
This paper surveys   mathematical models, structural results and  algorithms   in 
  controlled sensing with social learning in social networks.

Part 1, namely  {\em  Bayesian Social Learning with Controlled Sensing}, addresses the following  questions:
How does risk averse behaviour in social learning affect quickest change detection? How can information fusion be priced? How is the convergence rate 
of state estimation affected by social learning?
The aim is to develop and extend  structural results in stochastic control and Bayesian estimation to answer these questions. Such structural results yield fundamental bounds on the optimal performance, give insight into what parameters  affect the optimal policies,
and yield computationally efficient algorithms.

Part 2, namely, 
{\em Multi-agent Information Fusion with Behavioral Economics Constraints}, generalizes Part 1. The 
agents exhibit sophisticated decision making in a behavioral economics sense; namely the agents are \textit{rationally inattentive} (exhibit limited attention span) and make \textit{anticipatory decisions} (thus the decision strategies are time inconsistent and interpreted as  subgame Bayesian Nash equilibria). 

Part  3, namely {\em Interactive Sensing in Large Networks}, addresses the following questions:
 How to track the degree distribution of an infinite random graph with dynamics (via a stochastic approximation on a Hilbert space)? How can the infected degree distribution of a Markov modulated power law network and its mean field dynamics be tracked via Bayesian filtering
given incomplete  information obtained by sampling the network?
How does the structure of the network (Erd\H{o}s R\'enyi vs power law) affect  estimation of the infected
degree distribution and corresponding Cramer Rao bounds? 
We also briefly discuss how the glass ceiling effect emerges in social networks.

Part 4, namely \emph{Efficient Network Polling}  deals with polling in large scale social networks. In such networks,  only a fraction of nodes can be polled to determine their decisions. Which nodes should be polled to achieve a statistically accurate estimate of sociological  phenomena?  Some nodes may be reluctant to reveal their true opinion. 
  This may lead to incorrect polling estimates. How to compensate for this?

\end{abstract}

\end{titlepage}

\section{Introduction}
This paper discusses  controlled sensing and information fusion of interacting social sensors.\footnote{A social (human) sensor   provides information about its state (sentiment, social situation,  quality of product) to a social network  after
interaction  with other social sensors. In this paper, consistent with a large body of literature,  we adopt a  more stylized definition:  {\em a social sensor performs social learning}. One can   view a social sensor as an automated Bayesian decision system~(\ref{eq:myopic}).} 
In classical Bayesian signal processing, noisy observations  recorded by  physical sensors are used with Bayes rule to estimate an underlying state. 
Here we consider
controlled  sensing 
with  social learning to estimate the underlying state.
Social learning, or learning from the actions of others, is an integral part of human behavior and has been studied  in behavioral economics, sociology and 
 computer science
  to model the  interaction of  decision makers \citeB{Aum76,EF93,Ban92,BHW92,AO11,KT12,Cha04,EK10,KKT03,Say14b,WD16}.

 Social learning models  present unique challenges from a statistical signal processing point of view.
First,  social sensors \cite{CCL10,SOM10} interact with and influence each other.  For example, ratings posted on online reputation systems strongly influence the behavior of  individuals.
  This is usually not the case with physical sensors. 
Second, due to privacy
concerns and time-constraints,
social sensors  
 reveal  decisions 
(ratings, votes) which are  a quantized   function of their raw measurements and interactions with other social sensors. 
 
 \subsection{Background: What is Bayesian Social Learning? A Signal Processing Perspective}

To fix ideas, we start with a short review of \red{``vanilla" social learning}.
Let $\{\state_k\}$  denote a finite state Markov chain with state space $\statespace = \{1,2,\ldots,X\}$,
 transition matrix $\tp$, and initial distribution $\pi_0$.  Suppose at each time $k$,
 noisy measurements  $\obs_k \in \obspace$ are available, where $\obspace $ is a finite set and $y_k$ is generated from
 conditional distribution $P(\obs_k=y|\state_k = i)$.
 A multi-agent system aims to estimate the underlying state $\state_k$, at each time $k$.

In classical Bayesian estimation, the  multi-agent system has access to all previous observations. The  posterior $\belief_k(i) = \prob(\state_k=i| \obs_1,\ldots\obs_k)$ is computed  via 
the Hidden Markov model  (HMM) filter   \citeB{EAM95,EM02}\\
\begin{minipage}{5cm}
\red{Classical HMM Filter}
\end{minipage} \hspace{-2cm}
\begin{minipage}{13.4cm}
\beq \belief_{k+1} = \filter(\belief_k, \obs_{k+1}) \propto \underbrace{ \prob(\obs_{k+1}| \state_{k+1}=j)}_{\text{likelihood}}  \sum_i \tp_{ij} \, \underbrace{\belief_k(i)}_{\text{prior}}
\label{eq:hmmfilter} \eeq\end{minipage}

In sequential  social learning, agents do not have access to  observations of other agents.
Instead they only have access to the actions of previous agents.
Each agent acts once  in a predetermined sequential order indexed by $k=1,2,\ldots$.
Let  $a_k\in \{1,\ldots,A\}$ denote the action chosen by  agent $k$.
 Then  the posterior distribution  (public belief) $\belief_k(i) = \prob(\state_k=i| a_1,\ldots,a_k)$ is updated
via the following 3 step procedure \cite{Cha04}:\\
\begin{minipage}{2cm}
\red{Social Learning  Filter}
\end{minipage} 
\begin{minipage}{14.4cm}
\begin{align}
\priv_{k+1} &= \filter(\belief_k, \obs_{k+1})  \quad\;\quad\text{agent updates private belief using   HMM filter (\ref{eq:hmmfilter})} \nn \\
a_{k+1} &=  \argmin_a c_a^\p  \priv_{k+1}  \quad  \text{agent takes action $a_{k+1} = \argmin_a \E\{ c(x_{k+1},a) | a_{1},\ldots,a_{k}, \obs_{k+1}\}$} \label{eq:myopic} \\
{\belief_{k+1}} &=  \filtersocial( \belief_k, a_{k+1})   \propto  \underbrace{ \sum_{y\in \Y} \prob(a_{k+1}|y,\pi_k)\, \prob(y|x_{k+1}=j) }_{\text{likelihood}}\sum_{i}  \tp_{ij}\, \underbrace{\belief_k(i)}_{\text{prior}}  \quad  \text{(public belief update)} \label{eq:slf}
\end{align}
\end{minipage}\\
Note that action $a_k$ in (\ref{eq:myopic}) is a quantized version of  private belief $\eta_k$; this action  is made public.
The public belief $\belief_{k}$ computed by the social learning filter (\ref{eq:slf}) is the 
posterior distribution given all actions until time~$k$.

\red{\bf \em Key Point.} The social learning filter  (\ref{eq:slf}) has a remarkable structure:
{\em the likelihood is an explicit function of the prior } $\belief_k$; whereas in classical Bayesian filtering (\ref{eq:hmmfilter}), the likelihood  is
functionally independent of the prior.
This crucial difference results in unusual behaviour: herding, information cascades.\footnote{It is well known 
\citeB{Ban92,BHW92} that when 
the transition matrix $\tp = I$
(identity matrix),
the above social learning protocol  leads to an information cascade  in finite time
with probability~1.  The proof follows  from  the martingale convergence theorem.
Note that \begin{compactitem}
 \item  A {\em herd of agents} takes place at time $\bar{k}$, if the actions of all agents after time $\bar{k}$ are identical.
\item An {\em information cascade} occurs at time $\bar{k}$, if the public beliefs $\belief_k$ of all agents after time $\bar{k}$ are identical.
\end{compactitem}}
%
%
Social learning is used to explain why  customers choose crowded restaurants, and why financial booms and busts   occur \cite{Cha04}.

{\bf Perspective.}   Social learning originates in economics; yet it shares similarities with electrical engineering:
\begin{compactenum}
\item {\em Social Learning vs Decentralized Detection.}  Decentralized
detection (Tsitsiklis \cite{Tsi93}, Teneketzis \cite{TV84}, Varshney \cite{VV97}) falls within the class of {\em team decision theory} \cite{AT96} and shares
many  similarities to social learning but with key differences. Decentralized detection quantizes the observations, whereas social learning quantizes the Bayesian belief (\ref{eq:myopic}) leading to herd behavior and multi-threshold decision policies.
In decentralized detection the 
fusion rules are directly optimized where as in social learning the fusion rule  is prescribed (\ref{eq:slf}). 

\item  {\em Sequential Bayesian social learning.}
Part  1  considers controlled sensing  assuming {\em sequential interaction} between agents in a Bayesian social learning
framework. Even in this  simple sequential setup there are several open issues regarding controlled sensing. 
Social learning  
has been extended to more general graphs \citeB{AO11,AOT10,JMT13,KT13,RSV09,ADLO11,GJ12,Mue13}.

\item {\em Decision-enabled sensor networks.} The social learning framework also models   decision enabled sensor networks where each individual node is  a controlled  sensor. A natural question  is:  How to control
the interaction of autonomous decision makers as they learn from sensor data? Social learning with controlled sensing and fusion allows us to 
 achieve coordination in decision making.

\item  {\em Real Datasets.}  
Mathematical analysis alone cannot capture the complexity inherent in social sensing.
 See  \citeA{NHK16,HNK15,HKP20} for 
 analysis on YouTube   datasets.
   In \citeA{KH15,HSK14} we have studied herding  and misinformation propagation amongst human subjects with experimental psychologists;
   see also \cite{CK04}.
\end{compactenum}

\subsection{Context --  Three  Interesting Examples} \label{sec:context}

Controlled sensing  with social learning can result in interesting behaviour (at least to a statistical
signal processing audience).
Here are three examples.

\begin{figure}[h] \centering
\subfigure[Classical Quickest Detection has a convex (connected) stopping set]{\includegraphics[scale=0.09]{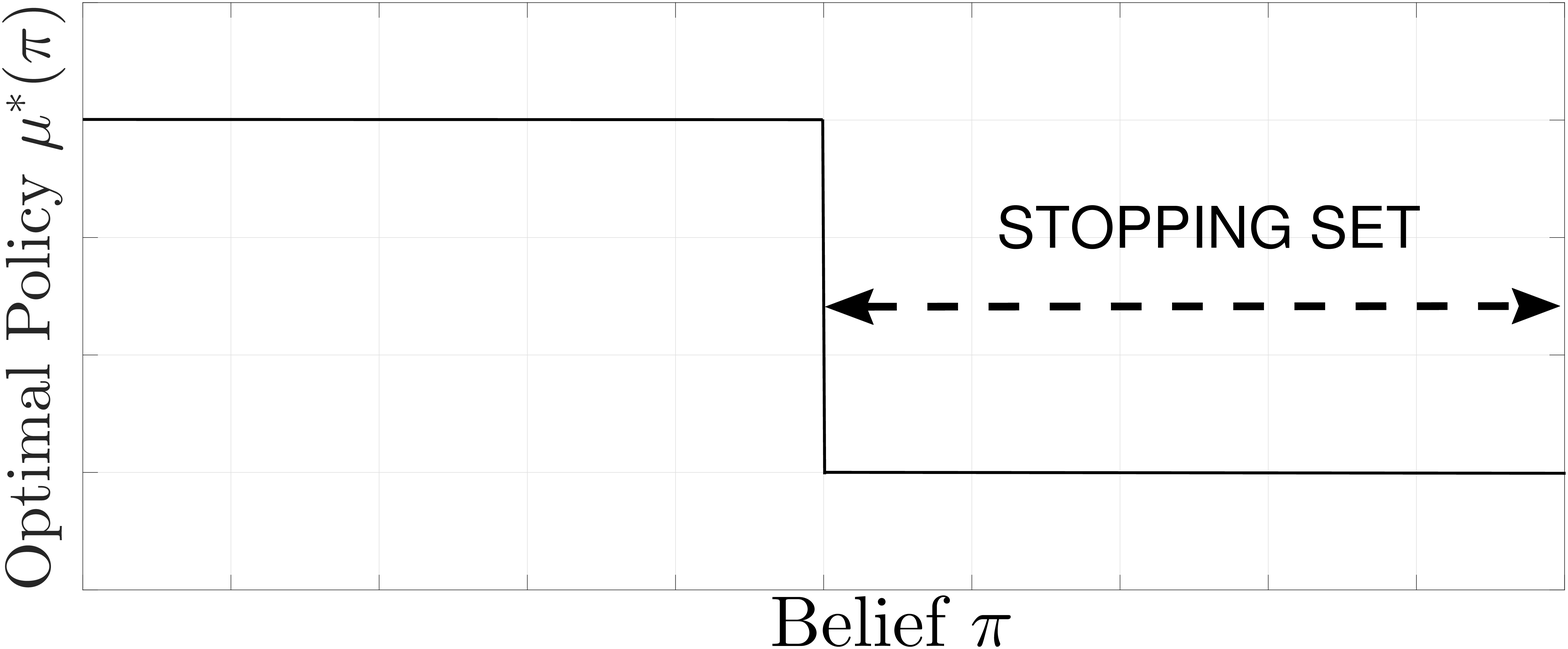}}    \hspace{18pt}
\subfigure[Quickest Detection with social learning has a disconnected stopping set]{\includegraphics[scale=0.09]{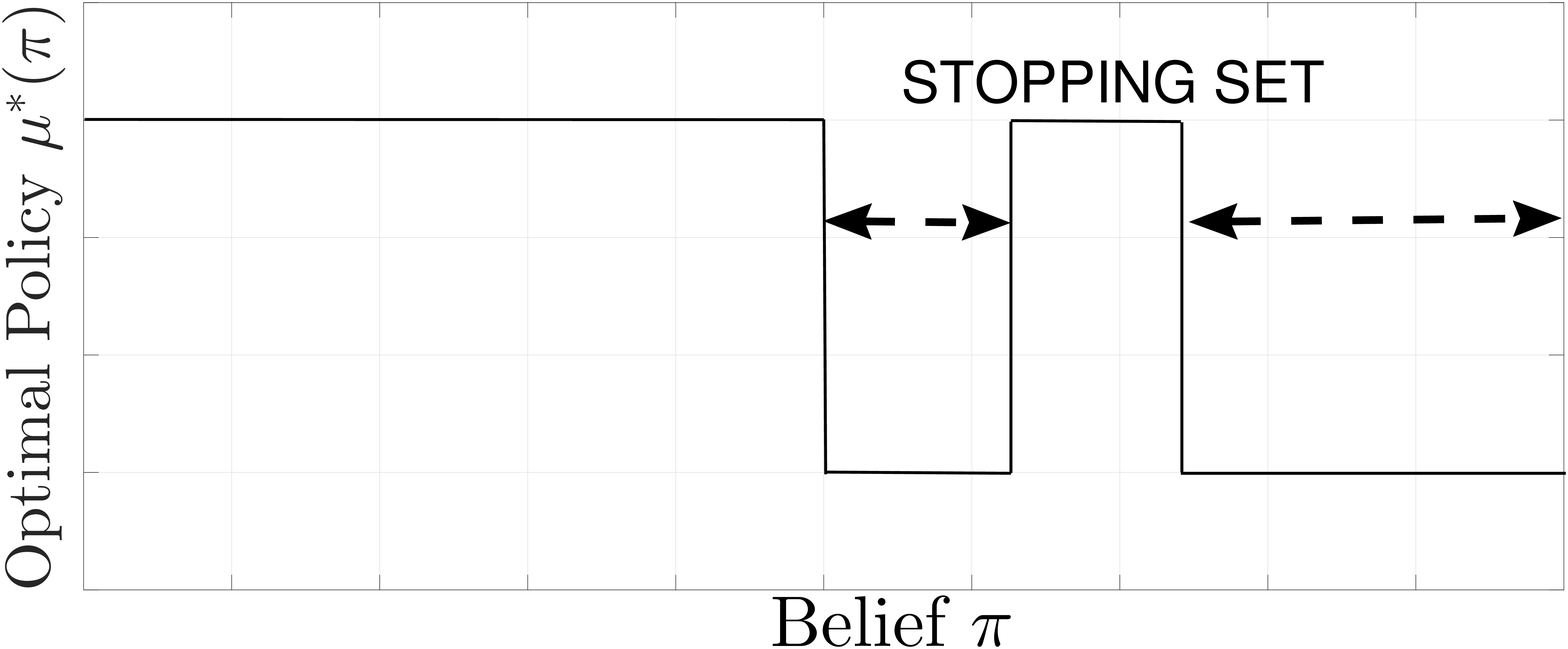}}
\caption[aa]{Change detection.  Classical vs  social learning - these results are  in our papers \protect\citeA{Kri12}, \citeA{KP15}.}
\label{fig:qdfig} \end{figure}

(i) {\bf \em Example 1. Change Detection with Social Learning can yield strange results}:
 Consider the classical Bayesian quickest change detection problem:  given noisy observations, detect if a  change has occurred in the underlying state   so as to minimize  a linear combination
of false alarm penalty and delay.
 It is well known \citeB{VB13,Shi63} that the  quickest  detection policy has a monotone threshold structure: when the 
posterior probability of change exceeds a  threshold, it is optimal to declare a change. Therefore, the optimal stopping set
(set of posteriors probabilities where it is optimal to declare ``change") is convex; see 
Figure \ref{fig:qdfig}(a).

Parts   1 and 2  generalize classical quickest detection to the case where    social sensors  interact via social learning (\ref{eq:myopic}), (\ref{eq:slf}).
  Based on the local actions generated by social learning in (\ref{eq:myopic}),  or equivalently, public belief $\belief$ computed by social learning filter (\ref{eq:slf}), a global decision maker performs quickest  change detection.
  
The optimal detection policy is shown in Fig.\ref{fig:qdfig}(b).
 The remarkable feature is    that  the stopping set is disconnected.  One sees
in Fig.\ref{fig:qdfig}(b)
 the counter-intuitive property:  {\em the optimal detection policy 
  switches from  announce ``change" to announce ``no  change" as  the posterior probability $\pi$ of a change increases!}

Thus   making a global decision as to  whether a change has occurred
based on local decisions of interacting agents is  non-trivial. 
This has implications in  anomaly/virality  detection in social networks \cite{SZY14,Moh14} and detecting market shocks in economics
\cite{PS11},\citeA{KB16}.
The result also has implications in automated sensing systems: be careful when quantizing Bayesian estimates (as in  (\ref{eq:myopic}))  and then performing
 change detection.

 The multi-threshold change detection  policy is a form of {\em change blindness}:  \textit{people fail to detect surprisingly
large changes to scenes} \cite{SR05}. A  human global decision maker might  ignore the multi-threshold optimal policy  and simply use the classical quickest detection policy. \citeA{Kri21} shows  that change-blindness is widely prevalent in anticipatory systems.

(ii) {\bf \em Example 2. Posterior Cramer Rao Bound (PCRLB)  for social sensing --  Power Law vs Erd\H{o}s R\'enyi:}

\begin{figure}[h]
     \centering  
     \includegraphics[scale=0.19]{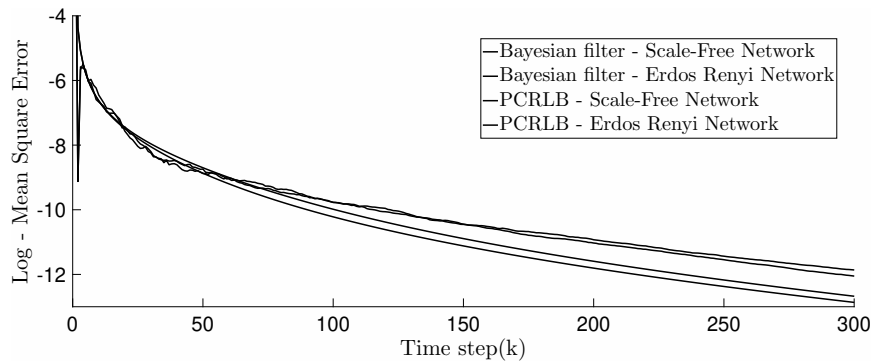}  
     \caption{Mean square error and PCRLB for tracking infected degree distribution for  Power law  and Erd\H{o}s-R\'enyi networks.  Both PCRLB and its slope are insensitive to the underlying distribution \protect\citeA{KBP17}.}
     \label{fig:pcmse}
\end{figure}
Part  2 considers social sensors that interact on large graphs.
We study  optimal  filtering to estimate  information propagation (infected degree distribution)  over time.
A natural question is: {\em How sensitive are the filtered estimates of the infected degree distribution to the underlying
graph structure?}  Two important graph structures are the    {\em power law} and {\em Erd\H{o}s R\'enyi} networks.
The power law network's degree distribution decays polynomially as $k^{-\gamma}$, $k=0,1,\ldots$ and occurs in
  online social networks like Twitter.
  The classical Erd\H{o}s R\'enyi  network's degree distribution  decays exponentially as $e^{-\gamma k}$.

The  PCRLB is a natural measure for the achievable performance of the mean square error of
the  filtered infected degree distribution estimate. 
Fig.\ref{fig:pcmse} compares the PCRLB for a power-law network 
versus an  Erd\H{o}s R\'enyi network. 
The surprising property in
Fig.~\ref{fig:pcmse} is   that  both the PCRLB and its slope are insensitive to the underlying network structure. 
This suggests that for tracking the infected degree distribution, precise knowledge of the underlying network distribution is not required.

(iii) {\bf {\em Glass Ceiling Effect   in Social Networks}}

\begin{wrapfigure}{r}{0.4\textwidth}
  \vspace{12pt} \centering
  \includegraphics[scale=0.45]{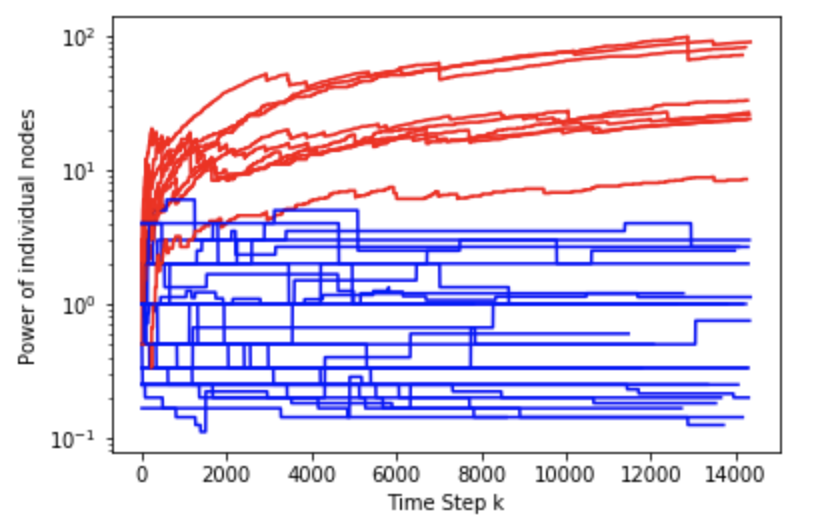}

  \mbox{}   \vspace{-0.8cm}
  
    \caption{Simulated glass ceiling effect in a time evolving  directed network (e.g.\ Twitter). The blue nodes constitute 80\% of the population but  their power never breaks through the glass ceiling. The power of a node  is the ratio of number of followers to number of followees.}
  \end{wrapfigure}
Here we illustrate a sociological dynamical effect in social networks that we will discuss in Part 3.    
 The \textit{glass ceiling effect} \cite{AKL15,ALP18}  refers
   to the barrier that keeps certain groups from rising to influential positions, regardless of their qualifications.
 In  analogy to real life,
   the glass ceiling effect is also highly visible in social networks such as Twitter, co-citation graphs and Instagram. In Twitter and Instagram, female users  have a smaller following compared to their male colleagues who are equally qualified \cite{SRC18,NGL16}. A similar empirical observation is found in co-citation graphs  where female authors receive less attention and fewer citations compared to their male colleagues
   and so are discouraged from academia \cite{WJK13,Iba97,SL05}. Our recent  work \citeA{NAI22,ANK22} shows that during  the global financial crisis in 2008, the glass ceiling effect on female authors was  more pronounced; possibly due to more female authors being furloughed/laid off, or forced to change jobs.
 Our aim is to develop and analyze dynamic network models 
   where  individual nodes 
   make decisions  leading to a minority of nodes  dominating high centrality positions in the network. This  will shed
   light on the interplay between sociological phenomena such as perception bias \citeA{ANA20}, homophily \cite{MSC01} and the  glass ceiling effect.

\section{Part 1. Controlled Sensing with Social Learning  -- A POMDP Approach}  \label{sec:cspomdp}
The main ideas revolve around two extensions of the vanilla social learning protocol (\ref{eq:myopic}),~(\ref{eq:slf}):
\begin{compactenum}
\item   Humans are  often risk averse decision makers \cite{TFT07} whereas  (\ref{eq:myopic}) specifies a social sensor
as an  expected utility maximizer.  How is social learning affected when risk  measures are used?
 Mathematically, a risk measure replaces the additive expectation operator in   (\ref{eq:myopic}) with a subadditive risk operator.
In  engineering and economics, risk averseness models   {\em robustness}; see  page \pageref{sec:dynamicrisk},  and also the paper  of  economics Nobel laureate Sargent \cite{HS01}.
Since social learning is  automated Bayesian decision making,  Part 1 can be viewed as  ``robust controlled sensing amongst 
automated Bayesian decision makers."
\item How does social learning interact with controlled sensing and controlled fusion? Social learning  (\ref{eq:myopic}), (\ref{eq:slf}) involves  exchanging myopic  decisions by social sensors  to estimate an underlying state; while controlled sensing involves stochastic control over a time horizon to estimate an underlying state.
This mismatch in time scales  between  myopic social sensors and a non-myopic controller (global
decision maker) results  in a non-standard controlled sensing problem.

\end{compactenum}

\subsection{POMDPs in Controlled Sensing:  4 Important Structural Results}
{\bf \em Why?} Partially observed Markov decision processes\footnote{For the reader unfamiliar with POMDPs: roughly, a  controlled-sensing POMDP is a  Hidden Markov model where the observation distribution $p(y|x,u)$ for any state $x$  can be
controlled with action $u$. The aim is to determine the optimal policy $\mu^*(\pi)$ which maps the belief (posterior) $\belief$ to the action $u$ in order to minimize an expected cumulative
cost over a (possible infinite) time horizon.}  (POMDPs) are a natural framework for sequential Bayesian decision making  under uncertainty. 
In this section   we summarize new and potentially very widely applicable results for  POMDPs that are of key importance in Part  1.
These results apply to wide variety of  problems  including quickest change detection, controlled sensing, and controlled fusion with social learning.

\begin{figure}[h] \centering
\includegraphics[scale=0.83]{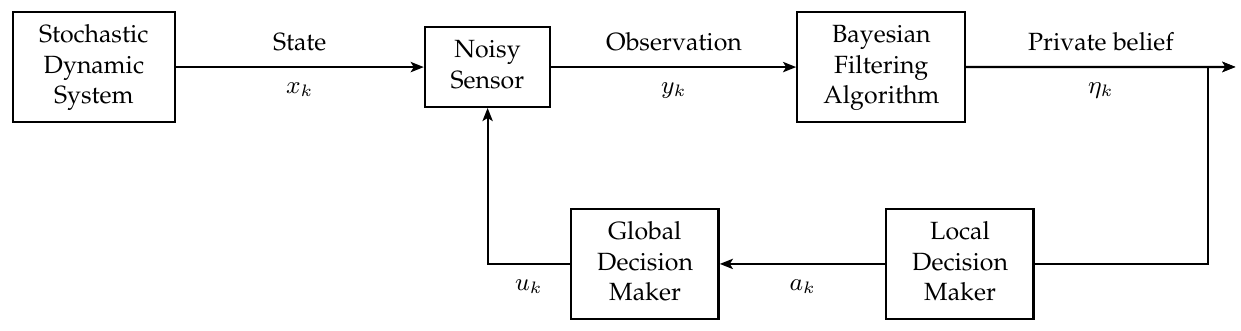}
\caption{Interaction of local decision makers (performing social learning) with a global decision maker that chooses decisions $u_k$ by solving a POMDP. We 
characterize the optimal policy in such controlled social sensing problems. The non-standard component compared to classical POMDPs is that  the controller (global decision maker) only has access to previous local decisions $a_k$ and not private beliefs.} \label{localglobal}
\end{figure}

{\bf \em Setup.} Figure \ref{localglobal} displays  our setup.  The key point is the interaction between local decision makers (social sensors) and the global decision maker:
 in quickest change detection, the global decision to continue or stop determines whether social learning is continued; and as the  local decisions accumulate via social learning, the global decision
maker must decide when to declare a~change.
This yields  a {\em non-standard POMDP} since the belief is updated using the social learning filter (\ref{eq:slf}) instead of  classical HMM filter (\ref{eq:hmmfilter}).
The optimal policy $\mu^*(\pi)$ satisfies
Bellman's dynamic programming equation, which  involves the social learning filter:
\beq  \label{eq:bellman}
 \optpolicy(\belief) = \argmin_{\action} Q(\belief,\action), \; V(\belief) = \min _{\action} Q(\belief,\action),  \text{ where } 
Q(\belief,\action) = C(\belief,u) +  \sum_a  V\big(\filtersocial(\belief,a,\action) \big) \, \sigma(\belief,a,\action) . 
\eeq
Here $\pi$ is the posterior (belief state) computed via the social learning filter  $\filtersocial(\cdot)$ in (\ref{eq:slf});  $C(\pi,u)$ is the cost incurred by the global decision
maker for choosing  action $u$ when the belief is $\pi$; and
$V(\pi)$ is the value function. Recall $a_k$ is the action of the local decision maker which is available to the global decision maker.

{\bf \em Why Structural results?}: Although POMDPs are a useful modeling paradigm for controlled sensing,
  computing  the optimal policy of a POMDP is in general intractable \citeB{PT87,Cas98,Lov91} since it requires solving Bellman's 
  equation (\ref{eq:bellman}) over the space of probability distributions (which is a continuum). We
 focus on 
 {\em structural results} for  the optimal policy rather than brute force numerical solutions. Structural results for POMDPs (studied in operations research, control theory
 and economics) yield  bounds on the optimal performance, give insight into what parameters  affect the optimal policies,
and yield computationally efficient algorithms. 
They involve deep results in lattice programming and stochastic orders.

Below are   four  essential structural results for POMDPs  (see 
\citeA{KD07,KD09,Kri11,Kri12,Kri16,KP15} and  \cite{Lov87,Rie91,RZ94} for  details).

\begin{thm}[Linear  Cost Stopping Time POMDP \cite{Lov87a}]  \label{thm:convex} Consider a stopping time POMDP with two actions: $u=1 \text{(stop)}$ and $u=2 \text{ (continue)} $ with  associated costs
$C(\pi,u)$.
 If $C(\pi,u)$ is linear in the belief $\belief$ (as  in standard POMDPs), then  the stopping set $\stopset$, namely 
$ \stopset = \{\belief:  \optpolicy(\belief) = 1 = \text{ stop}\}  $
is a convex set.
\end{thm}

Theorem \ref{thm:convex} is well known. It is the reason why classical quickest detection (with geometric or  more general phase-distributed change times) has a convex stopping region
\citeA{Kri11}.
Unfortunately, to model risk in social learning,  the cost  $C(\pi,u)$ is  non-linear in the belief $\pi$.  
Then the following novel generalization is required:

\begin{thm}[Threshold Optimal Policy for Non-linear Cost POMDP \citeA{Kri11}, \citeA{Kri16}] \label{thm:pomdpstructure}
If the (possibly nonlinear) cost $C(\pi,u)$ is increasing wrt $\belief $ in terms of first order stochastic dominance,
 then under reasonable conditions on the transition and observation probabilities:
 \begin{compactenum}
 \item The stopping set  
$ \stopset = \{\belief:  \optpolicy(\belief) = 1 = \text{ stop}\}  $
is a connected set (but possibly non-convex).  
\item The optimal policy $\mu^*(\pi)$ is increasing in $\pi$  with respect to the  monotone likelihood ratio stochastic order.
Therefore, $\mu^*(\pi)$ is characterized by a  single threshold curve that partitions the belief space. \end{compactenum}
\end{thm}

 \begin{wrapfigure}[9]{r}{0.3\textwidth}  
 \begin{center} \vspace{-0.5cm}
\includegraphics[scale=0.08]{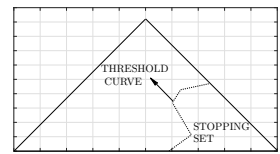}  \end{center} \vspace{-0.6cm}
\caption{Threshold  optimal policy for POMDP with nonlinear cost in  $\belief$. The stopping set is connected} 
 \label{fig:threshold} 
\end{wrapfigure}
 Theorem \ref{thm:pomdpstructure} 
 was developed in  \citeA{Kri11, Kri16}. It
   applies to nonstandard quickest
 detection with nonlinear penalties in the belief; such as variance and exponential costs \citeA{Kri11},\cite{Poo98}.
The last sentence in Theorem \ref{thm:pomdpstructure}  is the key point. Although computing the optimal policy via dynamic programming  is intractable, 
Theorem \ref{thm:pomdpstructure} says that under suitable conditions
 the optimal policy  is characterized by
a single threshold curve; see Fig.\ref{fig:threshold}. {\em One only  needs to estimate this threshold curve!} The optimal {\em linear} threshold policy\footnote{
This qualifies as  the {\em optimal  linear} policy
since we can give  conditions on the  coefficients of the linear threshold that are necessary and
sufficient for the resulting  policy to be  increasing  with respect to the  monotone likelihood ratio stochastic order \citeA{Kri11}.}
 which is increasing in $\pi$ can be computed straighforwardly via a simulation based stochastic gradient algorithm \cite{KY03,Spa03}.
Various parametrizations of the switching curve that result in a monotone policy can be obtained; such parametrized policies are optimal
within the given parametric class. 

 Theorem  \ref{thm:pomdpstructure} has two  underlying concepts.
 First, we first need to order belief states (probability vectors).  This is done using the monotone likelihood ratio order
or its multivariate generalization
called the TP2 (totally positive of order 2) order.
Despite being partial orders, these are ideal for POMDPs since they are preserved under Bayesian updates.

The second, important concept to establish the threshold curve in   Theorem \ref{thm:pomdpstructure} is  a novel form of submodularity\footnote{For the reader unfamiliar with this area,  submodularity and stochastic dominance on a lattice of belief states falls under the area of lattice programming. 
Lattice programming and monotone comparative statics are powerful tools pioneered by Topkis and now widely used in microeconomics \cite{Ath02,Ami05},
operations research \cite{SM02}, 
 control theory and game theory.} of Bellman's equation (\ref{eq:bellman}); novel because it applies to specific  line segments within the belief space.
These line segments form  chains, i.e., a totally ordered subsets for the likelihood ratio order.  
Submodularity implies $\mu^*(\pi) \uparrow \pi$ on each such  chain; the union of such chains covers the belief space
yielding the threshold curve.

The third result we outline below is 
  tight myopic
  upper and lower bounds for  the optimal policy  of  a POMDP. 
  \begin{thm}[Myopic Policy Bounds \citeA{KP15}]  \label{thm:myopic}
Under copositivity conditions on the  transition matrices, the optimal policy $\mu^*$ of a POMDP can be lower and upper bounded  by myopic polices
$\underline{\mu}$ and $\bar{\mu}$ so that
$$ \underline{\mu}(\pi) \leq \mu^*(\pi) \leq \bar{\mu}(\pi), \quad \text{ for all beliefs $\belief$ } $$
Moreover, choosing the cots associated with the myopic policies $\underline{\mu}$ and $\bar{\mu}$ to maximize the region $\pi$ where $\underline{\mu}(\pi) = \bar{\mu}(\pi)$ can be formulated
as a linear pogramming problem.

Under Blackwell dominance conditions \cite{Rie91} on the observation kernel (footnote \ref{foot:black}), $\mu^*(\pi)$   can be upper bounded by a myopic policy.
The tightest upper and lower bounds are obtained by solving a linear program.
\end{thm}
 Theorem \ref{thm:myopic} is useful for sub-optimal algorithms
and   establishing performance  bounds in controlled sensing.  They involve novel ideas in copositivity of stochastic
matrices  and Blackwell dominance \citeA{KP15,Kri16}.

Finally, we present a performance analysis result.
 Can  transition matrices $\tp$ and observation distributions $B$ be ordered so that the larger they are (with respect to a suitable partial order), the larger the optimal cumulative cost? Such a result is very useful -- it allows us to compare the optimal performance of different POMDP  social learning models, even though computing the optimal cost
or policy  is intractable. 
\begin{thm}[Performance Analysis  \citeA{Kri13,Kri16}] \label{thm:sens}
\begin{compactenum}
\item  Consider two distinct POMDP models $\model= (\tp,\oprob)$ and $\bmodel= (\btp,\boprob)$,
where $\tp \succeq \btp$ wrt copositive dominance and $B \succeq \bar{B}$ wrt Blackwell dominance.
Then  under reasonable conditions,  the optimal   costs   satisfy $J_{\mu^*(\th)}(\belief;\th) \leq J_{\mu^*(\bmodel)}(\belief;\bmodel). $
\item 
Consider two distinct POMDPs  $\model= (\tp,\oprob)$ and $\bmodel= (\btp,\boprob)$.
 Then for mis-specified model and mis-specified policy, the following sensitivity bounds hold (where constant $K$ can be determined explicitly)
\begin{align} \text{Mis-specified Model: } &
\sup_{\belief\in \I} |J_{\mu^*(\model)}(\belief;\model) - J_{\mu^*(\model)}(\belief;\bmodel)| \leq
K\|\model - \bmodel\| . \label{eq:sens0} \\
\text{Mis-specified policy: }  & J_{\mu^*(\bmodel)}(\belief,\model)  \leq J_{\mu^*(\model)}(\belief,\model) +  K \|\model - \bmodel\| \label{eq:sens1}. \end{align}
\end{compactenum}
 \end{thm}
 
Note that  (\ref{eq:sens1}) is a lower bound for the  cumulative cost of applying the optimal policy for a different
model $\bmodel$ to the true model $\model$ - this bound is in terms of the  cumulative cost of the optimal policy for true model $\bmodel$.
So  if the ``distance'' between the two models
$\model,\bmodel$  is small, then the performance loss  is small  (\ref{eq:sens0}), (\ref{eq:sens1}).

\subsection{Quickest Change Detection with Risk Averse Social Sensors} \label{sec:csqd}

Quickest detection is a simple example  of a
stopping time POMDP and serves as a useful example.
Recall the applications  in  virality detection and market shocks listed on page \pageref{fig:pcmse}.
(Quickest detection with measurement cost \cite{BV12},  quickest transient detection
\cite{PKV10}, quickest state detection are
other important examples; see \citeA{Kri11,Kri16} for formulation as a stopping time POMDP.)

In classical Bayesian quickest detection, the aim is to
determine the 
optimal policy $\mu^*$ to minimize the Kolmogorov--Shiryaev
criterion  \citeB{PH08,Shi63,VB13,TM10} which is a tradeoff between delay and false alarm penalty:
\beq J_\mu(\belief_0) =   d \,\Ep\{(\tau - \tau^0)^+ | \belief_0\} +  f \, \Ep\{ I(\tau < \tau^0 ) | \belief_0 \}.
\label{eq:ksd} \eeq
 Here $\tau_0$ is the change time, $\tau$ is the time when the detector announces a change has occurred, $d$ and $f$ denote  the delay  and false alarm penalties, respectively.
Classical quickest detection is a linear cost stopping time POMDP. Therefore, Theorem \ref{thm:convex} immediately implies that  the  optimal detection policy $\mu^*(\pi)$ 
  has a  threshold structure in the belief  $\pi$ (see Fig.\ref{fig:qdfig}(a)); where $\belief$ is computed by the 
HMM filter (\ref{eq:hmmfilter}).

Standard quickest detection (\ref{eq:ksd}) is  expectation centric.
We  consider two  generalizations:
\begin{compactenum}
\item Decision makers  influence each other via  the social learning protocol (\ref{eq:myopic}), (\ref{eq:slf}).  (In  signal processing, physical sensors usually  do not affect each other.)
So the belief $\belief$ is computed via social learning (\ref{eq:slf}). 

\item  Social sensors are risk averse. So the expectation operator $\E$ in social learning (\ref{eq:myopic}) and $\Ep$ in controlled sensing (\ref{eq:ksd}) 
are replaced by    {\em coherent risk measures}{\footnote{In simple terms, the idea is
to 
replace the  expectation operator which is additive with a more general {\em sub additive} risk measure.  Formally, a risk measure $\varrho : \mathcal{L} \rightarrow \mathbb{R}$ is a mapping from the space of measurable functions to the real line which satisfies the following properties: (i) $\varrho(0)=0$. (ii) If $S_{1}, S_{2} \in \mathcal{L}$ and $S_{1} \leq S_{2} ~\text{a.s}$ then $\varrho(S_{1}) \leq \varrho(S_{2})$. (iii) if $a\in\mathbb{R}$ and $S\in\mathcal{L}$, then $\varrho(S+a) = \varrho(S)+a $. The risk measure is {\bf coherent} if in addition $\varrho$ satisfies: (iv) If $S_{1}, S_{2} \in \mathcal{L}$, then $\varrho(S_{1}+S_{2}) \leq \varrho(S_{1}) + \varrho(S_{1})$. (v) If $a \geq 0$ and $S\in\mathcal{L}$, then $\varrho(aS)=a\varrho(S)$. The expectation operator is a special case where subadditivity is replaced by additivity. \label{foot:coherent}}}   \citeB{ADEH02,ADE07}; thereby
modeling risk averse local and global decision makers.
It is well documented in behavioral  economics \citeB{CLLS75,DS99} that humans prefer a certain outcome over an uncertain but potentially larger outcome. To model risk averse behaviour, widely  used coherent risk measures are  Conditional Value-at-Risk (CVaR) \cite{RU00,RU02}, Entropic/exponential  risk measure 
\cite{JBE94} and Tail value at risk \citeB{MJ10}; see also page \pageref{sec:dynamicrisk} for robustness interpretation of risk in terms of dynamic risk measures for the global
decision maker.
\end{compactenum}

\subsubsection{Why does Quickest Detection with Social Learning yield Strange Results?}

\begin{figure}
\centering
\includegraphics[scale=0.27]{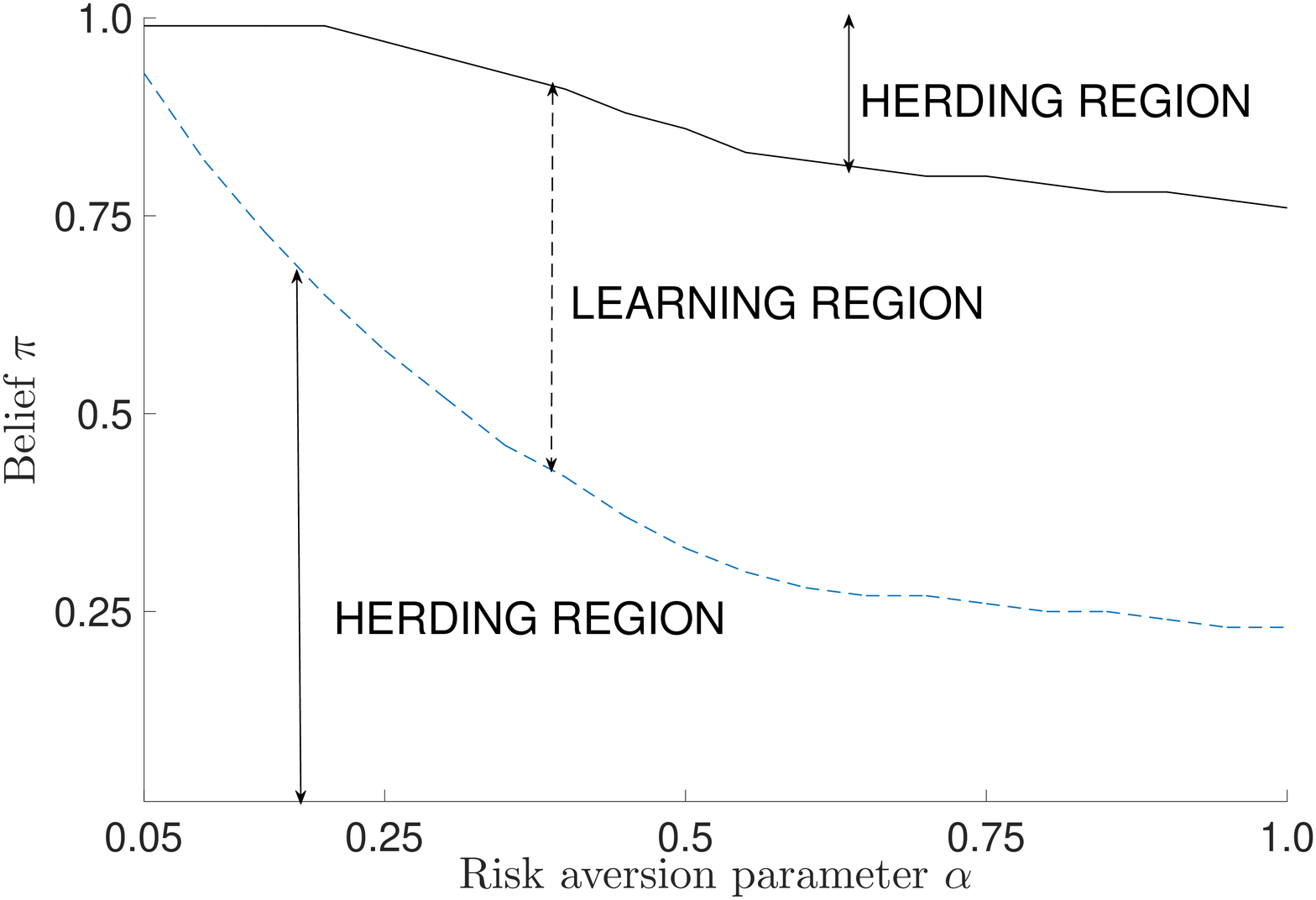} 
\caption{Effect of risk averse behavior on social learning \protect\citeA{KB16}. The figure shows the regions in belief space
where social learning and herding occur.
When social sensors are more risk
averse ($\alpha \rightarrow 0$), then the herding region  becomes larger. Social sensors   stick to the  decision of previous sensors
thereby exhibiting safety in numbers.} 
 \label{alp_p} 
\end{figure}
Consider the generalization of  classical quickest detection where a multiagent system performs social learning  \citeA{Kri12,KB16,Kri16}. 
Given the public belief  $\pi$ computed  by the social learning filter (\ref{eq:slf}), 
the global decision maker's policy $\mu^*:\pi \rightarrow \{\text{stop}, \text{ continue}\}$ that optimizes the quickest detection objective (\ref{eq:ksd}) 
satisfies
 Bellman's dynamic programming  equation for a stopping time POMDP; see (\ref{eq:bellman}) for notation: $\mu^*(\pi) = \argmin \{ \}$,
\begin{align} \label{eq:dp_alg}
 V(\pi) &= \min \{  f \pi(2), \; d(1-\pi(2))  +  \sum_{a \in \A}  V\left( \filtersocial(\pi,a) \right) \filterd(\pi,a)\}.
\end{align}

As discussed on page \pageref{sec:context}, the remarkable  result regarding Bellman's equation (\ref{eq:dp_alg}) is that the optimal policy is multi-threshold as shown in Fig.\ref{fig:qdfig}(b). This means that the optimal policy switches to ``no change" as the probability of change becomes larger!
Also the value function $V(\pi)$ is non-concave.  This is very different to a classical quickest detection  where the policy is monotone (threshold) and the value function is  concave. 

{\bf \em Unusual
 structure of 
 social learning filter.}  In a nutshell, the above unusual multi-threshold  behavior is due to 
the unusual structure of the social learning filter $\filtersocial(\pi,a)$  in (\ref{eq:dp_alg}). 
  For $P=I$, the social learning filter (\ref{eq:slf})   has two distinct behaviors depending on the prior belief $\pi$ \citeA{Kri12,Kri16}: 
  \begin{compactenum}
\item 
For $\pi$ in the herding region,  $\filtersocial(\pi,a) = \pi$. So the agent simply repeats the action of the previous agent. \item
 In the learning region,  $\filtersocial(\pi,a) = \filter(\pi,y)$, i.e., the social learning filter (\ref{eq:slf}) coincides with the classical Bayesian update (\ref{eq:hmmfilter}). 
 The size of the herding and learning regions depend on the costs $c(x,a)$ in (\ref{eq:myopic}).
 
 \end{compactenum}

As a result, of this discontinuity,    for two beliefs $\pi_1, \pi_2$ which are very close,
the updates can be vastly different. This results in a non-concave value function $V$ in (\ref{eq:dp_alg})  and hence,  a multi-threshold  optimal policy.

\subsubsection{Quickest Detection with Risk Averse Social Learning}
Next, consider the further generalization where expectation centric social learning is replaced with risk averse social learning, i.e.,  risk averse local decision makers.
To save space,  we focus on  the  CVaR risk measure\footnote{
For the reader unfamiliar with risk measures,  CVaR is one of the `big' developments in risk modelling \citeB{RU00,RU02}.
In comparison, the 
 value at risk (VaR) is the percentile loss namely,  $\text{VaR}_\alpha(x) = \min\{z: F_x(z) \geq \alpha\} $ for cdf $F_x$. While CVaR is a coherent risk measure,
 VaR is not  convex and so not coherent. CVaR has   remarkable properties: it is continuous in $\alpha$
 and jointly convex in $(x,\alpha)$.
For continuous cdf $F_x$, $\text{CVaR}_\alpha(x) = \E\{X | X > \text{VaR}_\alpha(x)\}$. Note that the variance  is
 not  a coherent risk~measure.}; however, our research will study general risk averse measures. In analogy to social learning (\ref{eq:myopic}),
given  observation $y_k$,
the  local decision maker  chooses  action $a_{k}$ at time $k$ to minimize the CVaR cost:
\beq
a_{k} =  {\underset{a }{\text{argmin}}} \{ \cvr \} 
=  {\underset{a}{\text{argmin}}} \{ {\underset{z \in \mathbb{R}}{\text{min}}} ~ \{ z + \frac{1}{\alpha} \mathbb{E}_{y_{k}}[{\max} \{ (c(x_{k},a)-z),0 \rbrace] \} \} . \label{eq:riskmyopic}\eeq
Here $\alpha \in (0,1]$ is  the degree of risk-aversion for the agent (smaller $\alpha$ implies  more risk-averse behavior).
(\ref{eq:riskmyopic}) together with (\ref{eq:slf}) constitutes the CVaR social learning filter.

{\em Q1. What are the structural properties of the risk averse social learning filter?}
 \citeA{KH15} shows that, under reasonable assumptions on the costs, the decisions taken by risk-averse agents are ordinal functions of their private observations and monotone in the prior information.
  Thus Bayesian social learning follows simple intuitive rules and is a  useful idealization of 
human behavior;  see the highly influential paper \cite{Mil81}.   
 Fig.\ref{alp_p} shows  the herding regions in the belief space for CVaR risk averse
social learning filter.  It can be observed from Fig.\ref{alp_p} that the region of beliefs 
  where social learning occurs  grows smaller  as the  parameter $\alpha$ decreases (agents become more risk
averse).  
 So Fig. \ref{alp_p} can be interpreted as saying that  risk-averse agents show a larger tendency to go with the crowd rather than ``risk" choosing the other action. 
 In particular as $\alpha\rightarrow 0$ (extreme risk averse case), the entire state space becomes a herding region.



{\em Q2. How is Quickest Change Detection Policy affected by risk averse social sensors?}  
 \citeA{KB16} shows  numerically that the stopping region is  non-convex. Our objective is  to characterize  the optimal policy (extensions of Theorems \ref{thm:pomdpstructure}, \ref{thm:myopic}, and \ref{thm:sens}).

{\em Interpretation}:
Multi-threshold behavior reflects the  lack of confidence by the global decision maker: if it is optimal to announce  a change, 
it may not be optimal to declare a change when the belief is higher. 
So the above conjecture says:  If the decision maker is confident of announcing a change for a certain risk aversion, then it remains confident 
of its decision when the agents are more risk averse (more careful).
Another interpretation is in terms of change blindness as discussed in the Introduction.

 {\em Q3.  Optimal Achievable Cost?} The risk averse behavior in social learning manifests itself in the action likelihood (\ref{eq:slf}).  This likelihood is the product
 of 
 the classical likelihood matrix with a conditional probability. We will use Blackwell dominance\footnote{A stochastic  kernel $B$ Blackwell dominates  $\bar{B}$ if $\bar{B} = B Q$ for some stochastic kernel $Q$. Put simply, $\bar{B}$ is more noisy than $B$.
\label{foot:black}} in Theorem   \vref{thm:sens}  to show that
the cost incurred by quickest detection with risk averse social learning is always larger than classical quickest detection.
Also Theorem \vref{thm:myopic} yields  myopic bounds that sandwich the optimal policy and upper bounds to the achievable
cost.



\subsubsection{Robust Quickest Detection with social learning using dynamic risk measures} \label{sec:dynamicrisk}
Thus far, we have considered quickest detection where  {\em risk averse local decision} makers that perform social learning. Now we consider a {\em risk averse global decision} maker that performs
quickest change detection. Since the global decision maker solves a POMDP, we need to use {\em dynamic risk measures}.

{\bf \em Why?} Dynamic  risk averse measures  model {\em robustness} in decision making.
Let us quickly explain this.  The  Kolmogorov-Shiryaev criterion (\ref{eq:ksd}) for quickest detection optimizes the additive  objective  $ \Ep\{J\}$ wrt policy $\mu$,
 where $J  $ is the accumulated  sample path cost.
 In risk averse control with an exponential penalty, one seeks to optimize $J_R = \Ep\{\th \exp(\th J) \}$ where $\th >0$ denotes the risk averse parameter. Expanding the exponential yields
$ J_R = \th + \th^2 \Ep\{J\} + \th^3 c_\th  \Ep\{J^2\}$ for some constant $c_\th$.
So   $J_R$  penalizes
 heavily large sample path costs due to the presence of second   order moments and therefore is used widely in robust  control \cite{Whi81b,PJD00,Ben92,HS01}.
  Interestingly,
 \cite{Poo98} published in 1998 is the first  exponential risk quickest detection paper.
Of course, today the exponential penalty is just one of a large number of risk averse measures. In particular, with the  rapid recent progress in  {\em coherent} risk measures
 (see footnote \ref{foot:coherent}), there is strong
 motivation to develop robust controlled sensing  results; see also \cite{UVM11} for minimax robustness.

{\bf \em How?}
Risk averse control replaces
the  expectations   $\E_\bpolicy\{ \cdot | \belief_0\}$  in stochastic control 
by more general subadditive {\em dynamic} risk measures.
Accordingly,
  consider the global decision maker's   risk averse   objective 
\beq
 J_\bpolicy(\belief_0) = \riskf ^0_\bpolicy\biggl\{ c(\state_0,\action_0) +  \riskf ^1_\bpolicy \bigl \{ c(\state_1,\action_1) + \cdots + 
 \riskf ^{\finaltime-1}_\bpolicy \{\cost(\state_{\finaltime-1},\action_{\finaltime-1}) + \cost_\finaltime(\state_{\finaltime} ) \} \bigr \}  \vert \belief_0 \biggr\}.
 \label{eq:robj}
\eeq
Here, the one step risk measures  $\riskf ^0_\bpolicy, \riskf ^1_\bpolicy, \ldots$ replace the expectation operator $\E_\bpolicy$
in  (\ref{eq:ksd}).
The operators $ \riskf ^k_\bpolicy\big\{ \fun(\state_k,\action_k, \state_{k+1}) \big\}$ are called  {\em Markov risk transition mappings} and are   a sub-additive generalization of the
conditional expectation $\E\{\cdot | \state_k,\action_k\}$, see \cite{Rus10,CR14} for detailed exposition of dynamic risk measures.
%
%

Bellman's equation for quickest detection with dynamic CVaR risk  and social learning reads   (recall from (\ref{eq:riskmyopic}) that $\alpha \in (0,1]$ models the degree of risk-aversion):
optimal policy $\mu^*(\pi) = \argmin\{\cdot\}$ and
 \begin{align} \label{eq:belrisk}
 V(\belief) = \min\{ f \, \pi(2) ,  d \, (1 - \pi(2)) + \inf_{z \in \reals} \big\{ z + \frac{1}{\alpha} \sum_y \max\{ V(\filtersocial(\belief,a)) - z, 0 \} \filterd(\belief,a) \}.
 %
 %
\end{align}
Notice the substantial difference compared to  the risk neutral Bellman's equation (\ref{eq:dp_alg}) for quickest detection.
 In the most general setup we will consider: individual social sensors  are risk averse in their
 decision making, and a dynamic risk measure
is used in the control objective for the global decision maker. We can constrict structural results for the optimal policy by extending 
Theorem \vref{thm:pomdpstructure}, see
 \citeA{Kri16,KB16}.  Using Theorem \vref{thm:myopic} we can construct myopic policies that provably upper and lower bound the optimal policy.
 This  allow us to determine achievable   bounds to the optimal cost (see Theorem \vref{thm:sens}).

\subsection{Controlled Information Fusion with Social Sensors} \label{sec:cif}

Thus far,  the interaction of global and local decision makers was limited to  controlling the  local decisions over time.   We now consider 
controlled social learning.
 We wish
to determine \begin{quoting} How  to price the quality of local decisions over time, to  achieve {\em controlled
information fusion?} \end{quoting}

In the classical social learning protocol (\ref{eq:myopic}), (\ref{eq:slf}), 
social sensors  are interested in minimizing  their own myopic costs~(\ref{eq:myopic})  and   ignore the information benefits their action provides
to others.
This leads to an information cascade and social learning stops (when estimating a random variable).
Our aim is to devise algorithms to delay herding until the state estimate has reached acceptable accuracy. Below we discuss how the global decision maker can affect the quality of the observations (actions) amongst  the social sensors in three  ways: 
\begin{compactenum}
  \item 
Controlling the information structure (Sec.\ref{sec:privacy})  
\item  Pricing information fusion (Sec.\ref{sec:optprice})  
\end{compactenum}
These   constitute {\em socialistic learning}:  agents coordinate their actions to achieve a common goal.

\subsubsection{Controlling the Information Structure in Social Learning -- Privacy vs Reputation} \label{sec:privacy}

{\bf \em Why?} Controlling the  information structure is important in adaptive trust/reputation social systems \cite{KTA05,MMH02}. Can  social sensors  assist social learning by choosing their actions to trade off individual privacy  (local costs) 
while  optimizing the reputation of their  social group
(global cost)?
 Building on \citeA{Kri12,Kri13},
agents  minimize the  welfare cost involving {\em all} agents  $k=1,2,\ldots$ in their group  (compared to
myopic objective~(\ref{eq:myopic})):
\beq \label{eq:pomdp}
J_\mu(\belief_0) = \Ep \biggl\{\sum_{k=1}^\infty \discount^{k-1} c(x,a_k)  
| \pi_0  \biggr\}  ,
\; \text{ subject to privacy constrained  rule }   a_k = a(\pi_{k-1},y_k,\mu(\pi_{k-1})).
 \eeq


The key point in (\ref{eq:pomdp})  is that each agent  chooses its  action according to the
{\em privacy constrained rule} \cite{Cha04,SS00}.    The strategy  $\mu: \pi_{k-1} \rightarrow \{1,2\ldots, L\}  $
maps the available public belief to the set of $L$ privacy values.
The higher the privacy value, the less the agent reveals through its action.  
This is  in contrast  to classical social
learning  (\ref{eq:myopic}) where  the action $a(\pi,y)$ is chosen as a myopic function of the  observation and public
belief.

{\bf \em How? Quickest Herding optimal policy}.
The   optimal policy $\mu^*(\belief)$  that 
minimizes  (\ref{eq:pomdp})  satisfies Bellman's equation    
 (\ref{eq:bellman}).

To gain  insight,
suppose there are two privacy values  and each agent $k$ chooses action  
  $$ a_k = \begin{cases} y_k  & \text{ if } \mu(\pi_k) = 1 \text{ (no privacy)} \\
  \arg\min_a c_a^\p \pi_{k-1} & \text{ if }  \mu(\pi_k) = 2 \text{ (full privacy)}.
   						   \end{cases}  
$$  
So an agent either reveals its observation (no privacy) or chooses its action by  neglecting its observation (full privacy).
 Notice that once an  agent chooses full privacy, then all subsequent agents  choose
 the same action and therefore herd - this follows since each agent's
  action reveals
  nothing about the underlying state. 
  Therefore, the problem is a stopping time POMDP:
Determine  the earliest time for agents to herd (maintain full privacy) subject to maximizing the
social group reputation.  We have the following results:
\begin{compactitem}
\item
The classical Theorem \ref{thm:convex} does not apply since the costs $\argmin_a c_a^\p \pi$ are non-linear in belief $\pi$. 
\item But
  Theorem \vref{thm:pomdpstructure} from our previous work \citeA{Kri11,Kri16} does apply! It says that 
 the optimal policy $\mu^*$  has a threshold structure.  Indeed the stopping set is non-convex but connected! 
 \item 
If individuals deploy the heuristic {\em ``Choose increased privacy when belief is close to the target state,''} then  group behavior
is sophisticated:  herding is delayed
 and accurate state estimates   are obtained. \end{compactitem}

{\em 
How can the quickest herding results be extended to risk averse agents and dynamic risk measures? }  As in Theorem \vref{thm:pomdpstructure}, the stopping region is connected but non-convex. This is in stark contrast  to  the results of the previous
section where the stopping region is the union of disconnected sets.
With Theorem \ref{thm:myopic}, the optimal policy can be sandwiched by myopic lower and upper bounds; and the region where these bounds overlap can be
optimized via a linear program.

\subsubsection{Optimal Pricing of Information Fusion in Social Learning Framework} \label{sec:optprice}
We next turn to  controlling fusion by pricing information; which is important in automated decision systems \cite{BBL12}.
 Suppose  a fusion center pays social sensors to choose actions that help the fusion
center estimate the underlying state.  For $k=1,2,\ldots$, let $p_k$ denote the price paid to social sensor  $k$ to choose an action that reveals
its private observation~$y_k$. So the cumulative payment (cost) accrued by the fusion center is 
\beq  J_\mu(\belief) = \E_\policy\{ \sum_{k=1}^\infty \discount^{k-1}\monprice_{k} \, I(a_{k}= y_k) \mid \belief_0 = \belief \}, \quad
\text{ where } \monprice_k= \policy(\belief_{k-1}). \label{eq:pricingfuse}
 \eeq
 Here $\pi_k$ is the belief state computed via the social learning protocol (\ref{eq:slf}).
{\em How can the fusion center minimize its cumulative payment (\ref{eq:pricingfuse}) while simultaneously maximizing the accuracy of the state estimate?}

\citeA{BK20} reveals two interesting results:
\begin{compactenum}
\item  The optimal price sequence is a {\em supermartingale},  that is, $\E\{p_{k+1} | a_1,\ldots,a_k\}\leq  p_{k} $ implying that 
 $\E\{p_{k+1} \}\leq \E\{ p_{k} \}$. That is, on the average, it is always better to   pay agents more to obtain more accurate estimates
initially,
and subsequently, as the state estimate gets more accurate, pay less. 
\item The optimal pricing policy  $\mu^*(\pi)$ has a piecewise monotone structure in the belief state $\belief$. This follows from Theorem \vref{thm:pomdpstructure}.
\end{compactenum}

\paragraph{Monopolist Interpretation.} 
The above result has a useful parallel in economic pricing  \cite{Cha04} that we will explore and generalize.
%
%
Suppose instead of $I(a_k=y_k)$ in (\ref{eq:pricingfuse}) one uses $I(a_k=1)$ which denotes a customer buying a 
 product from a monopolist at price $p_k$.
Each time a  customer buys from the monopoly, two things happen: the monopoly makes money due to the sale; also its gets publicity
 via social learning (review on social media).
 The monopolist chooses its price $\monprice_k= \policy(\belief_{k-1})$   based on the public belief $\belief_{k-1}$ 
 to
 maximize its cumulative discounted reward.
The optimal price  is again a supermartingale \cite{Cha04}: so    $\E\{p_{k+1} \} \leq \E\{p_k\}$.\footnote{One might conjecture that the monopolist  starts at a low
promotional price and then increase prices - but under the assumptions here, that is not optimal. The optimal price is a supermartingale:
the expected optimal price starts  high and then gets lower. In real life,  
it can be argued that 
elite companies such as Apple and Tesla often start at a high price to establish an elite customer base.}  

\paragraph{Summary} Part  1 dealt with Bayesian estimation and controlled sensing/fusion involving sequential social learning.  The main ideas 
involved structural results (stochastic orders and lattice programming)  for risk averse controlled
sensing of social sensors and pricing information fusion in POMDPs.

\section{Part 2. Multi-agent Information Fusion. Behavioral Economics Constraints} 

We now discuss   multi-agent information fusion with
behavioral economics constraints.
This  is a substantial generalization of classical
data fusion. In classical  data fusion, a fusion center
combines  estimates from  {\em physical sensors} to estimate an underlying state.
Even in its classical form, information fusion with social sensors is challenging due to inefficiencies in social learning  \cite{BHW92,Ban92,Cha04,CK04,MTJ18,ZCP13} such as herding and information cascades, i.e.\,, agents ignore their own observations and parrot decisions of previous agents.
Moreover, having more social sensors is not always advantageous (in  reducing mean square error) --  crowds can be more biased  than individuals. 

Part 2 generalizes Part 1 as we 
 consider  multi-agent information fusion when individual agents are sophisticated decision makers in a behavioral economics sense - they exhibit rational inattention and are anticipatory. 
Unfortunately, classical social learning is too simplistic to model the peculiarities of human decision making.  Adding   behavioral economics constraints to social sensors
  presents unique challenges from a statistical signal processing point of view.
First,  agents  have limited attention spans. 
According to the paradigm of \emph{rational inattention}\footnote{Rational inattention is a form of \textit{bounded rationality}. To quote  \cite{Sim10}: ``Limits on attention impact choice. For example, purchasers limit their attention to a relatively small
number of websites when buying over the internet; shoppers buy expensive products due to their failure to notice if sales tax is included in the price.''},  attention is a time-limited resource that can be modeled in terms of an information-theoretic (Shannon) channel capacity.
In statistical signal processing  terms, rational inattention is a form of controlled sensing: obtaining a more accurate observation requires more attention by the social sensor and therefore costs more.

Second,  agents are   \textit{anticipatory}: they make decisions by taking into account
   the probability of future decisions (plans).
In anticipatory decision making, the dependence of the current
reward on future plans results in a deviation between planning and execution. This 
phenomenon is called \textit{time-inconsistency} \cite{BM14} and 
 Bellman's principle of optimality no longer holds.  The appropriate formalism is the subgame Nash equilibrium. In game-theoretic terms, time-inconsistency arises when the optimal
 policy to the current multi-stage decision  problem is sub-game imperfect.
 Anticipatory decision making is studied extensively in behavioral
economics \cite{CL01,BPP17} because it  mimics important features of human decision
making.
Studies in psychology, neuroscience
\cite{BT16,CBS18,BP05,BPP17} indicate  that humans are anticipation-driven, and even
simple decisions  involve  multi-stage planning.  Time inconsistency of anticipatory decision  making  results in the 
{\em planning fallacy} of  Kahneman \& Tversky~\cite{KT79}: people tend to be optimistic and underestimate the time required to complete a future task.
It is therefore important  to  incorporate behavioral economics constraints to model and analyze interacting social sensors.

{\em Remark}.
The reader may be familiar with  \textit{risk-sensitive/averse} stochastic control \cite{JBE94,GM97}. Traditionally, risk aversion is a \textit{static} concept whereas
anticipatory decision making involves a multi-stage setting.
For example, CVaR (conditional value-at-risk \cite{RU00,RU02}) is a static coherent risk measure.
This distinction blurs in more general {\em dynamic} risk formulations \cite{Rus10,CR14}, but then the clarity and elegant structure of human  anticipatory decision making is lost; see also 
\citeA{KB16,KAB18}.

\subsection*{Background. Anticipatory Agents, Rational Inattention, Time Inconsistent Decision Making} 
Anticipatory decision making and rational inattention are  two  important behavioral economics models;  we briefly discuss them in a simplified mathematical sense to explain our ideas.

{\bf Anticipatory Decision Making}. The key idea behind the anticipatory decision model \cite{CL01} is as follows:  Let  $s_1,s_2$ denote an underlying state at times 1,2.
Then an anticipatory agent  chooses  decisions $a_1, a_2$ from strategies
$\mu_1,\mu_2$ to
maximize a    2-stage     utility (\cite{BPP17} has a  general multi-stage
formulation)
 \begin{align}
   \sup_{\pol_1,\pol_2}  \utilitytogo(s_1,{\pol_1,\pol_2}) &=  \sup_{\pol_1,\pol_2} \E_{\pol_1,\pol_2}\{\reward_1(s_1,a_1,\underbrace{\{p(a_2=a|s_1,a_1,\mu_2),a\in A\}}_{\text{anticipatory term}}) + 
                                              \reward_2(s_2,a_2) \}  \label{eq:cl_cost}
 \end{align}
 The 2-stage  utility  (\ref{eq:cl_cost})  looks just like a standard time separable utility for a Markov decision process, except for the \textit{anticipatory term}  $\{\pdf(a_2=a|\physical_1,a_1,\pol_2),a\in {A}\}$  in the reward  $\reward_1$;
 this models anxiety (psychological state) of  the decision  maker.
 The dependency of the reward at time 1  on the probability of future actions  (at time 2)
 results in \emph{time inconsistent} decision making\footnote{\cite{BM14} classifies time inconsistent decision problems into 3 types: non-geometric discount factor, problems where the cost depends on the reward depends on future state and  action probabilities (our framework), and nonlinear terminal cost problems.}; Bellman's  principle of optimality does not hold \cite{ACD14,BM14}.
 
The  time inconsistent decision problem (\ref{eq:cl_cost}) is `solved'
using the so called \textit{extended Bellman equation} \cite{CL01,BM14} to obtain the subgame perfect Nash equilibrium strategies
\beq 
 \mu_2^*(s_2) = \argmax_{a_2} r_2(s_2,a_2), \quad
\mu_1^*(s_1) = \argmax_{a_1} J(s_1,a_1,\mu_2^*)  \label{eq:BNE} \eeq
For  time inconsistent problems, neither the Nash equilibrium $\mu_1^*,\mu_2^*$   nor its value  $J$
are unique \cite{BM14}.  This is in contrast to classical  dynamic programming  where the optimal value is always unique.
Note that the subgame Nash equilibrium approach to time inconsistency
disregards
the fact that $\pol_2^*$ is no longer optimal at time 1. Another insightful way of viewing this is  that
 the   estimated  anticipatory
reward $\reward_1(\cdot)$ requires the agent to extrapolate what might happen at the second
stage, plans are not optimal once an action is taken \cite{BM14}. 

Obviously humans  do not solve 
time inconsistent  problems  to make decisions;  even professors struggle with such formalisms! Time inconsistent
behavioral economics models are widely used  because  they provide {\em generative} models for   the peculiarities of    human decision making. Indeed
\cite{CL01} shows that the above simple two-stage model
captures  important aspects of time-inconsistent human decision making: 
\begin{compactenum} \item
 Anticipatory agents act to reduce anxiety.  \cite{CB64}
presents experimental results where  people chose a more painful electric
shock today than waiting anxiously for a less painful  shock tomorrow.  
\item  Anticipatory agents deliberately avoid
  information by `sticking  their head in the sand' (with obvious consequences in  decision making). \cite{MM83} reports that giving patients
  more information
of a medical surgery procedure raised their anxiety and decreased
their probability of choosing the correct decision.
\end{compactenum}

{\bf Multi-agent Rational Inattention}. Finally a few words about rational inattention 
Sims \cite{Sim03,Sim10}.
To quote  Sims
``\textit{Rational inattention  models introduce the idea that
people's abilities to translate external data into action are constrained by a finite
Shannon capacity to process information}''.

A  \emph{multi-agent} 
rational inattention Bayesian model has the following dynamics  \cite{CD13,CD15,SRM16,CDL19}:
Given decisions $a_1,\ldots,a_{k-1}$ of previous agents,  agent  $k$ chooses its observation $\obs_k$ with \textit{attention span} $u_k$ and then makes decision $\laction_k$ via the following two-stage optimization
(where cost $c$  depends  on an underlying state $x$)
 \begin{equation}
   \begin{split} (\action_k, \laction_k) &= \argmin_\action \E_\obs \big\{ \argmin_{\laction} \underbrace{\E\{ \cost(\state,\laction) | \laction_{1:k-1},\obs_k,\action\}}_{\text{expected cost}} +
     \lambda \big[ \underbrace{H({\belief}) - H( {\filter(\belief,y,u)})}_{\text{ {rational inattention cost}}} \big]\big\}  \end{split}
   \label{eq:ra} \end{equation}
 Here $\pi$ is the prior, and 
 {$\filter(\pi,y,u)$ denotes the Bayesian posterior} given observation  $y$ and attention  $u$. Also
 $\lambda$ is a positive constant 
 and  $ H(\belief) = - \sum_{i=1}^\statedim \belief(i) \log_2 \belief(i) $ is the Shannon entropy.

The  mutual information between the prior and posterior, namely,  $H(\belief) - H(\filter(\belief,y,u) )$ in (\ref{eq:ra}), is called the \textit{information acquisition cost} in behavioral economics.
 Clearly, choosing a higher attention  span yields a more accurate  observation $\obs$ of  state $\state$ and  reduces  uncertainty in the  updated belief $\filter(\belief,y,u)$; so the entropy $H(\filter(\belief,y,u))$ becomes smaller.
 Thus, the information acquisition  cost becomes larger. So
 rational inattention
 (\ref{eq:ra}) trades off  observation accuracy with information acquisition cost.
 
Obviously  humans do not solve a Bayesian optimization
  with  entropy cost to make  decisions! Rational inattention  it is a useful  {\em generative} model. Experimental studies \cite{Sim03,Sim10,CD13,CD15} show that human decision making is consistent with rational inattention.  Our recent work \citeA{HKP20} analyzes a massive YouTube data set and shows that the commenting behavior of users is  consistent with
  rational inattention;
  indeed these 
  models yield remarkably accurate  predictive performance.

\subsection{Multi-agent Quickest  Detection with Anticipatory Decision Makers} \label{sec:csqd}

With the above background, we now discuss multi-agent quickest detection.
We study quickest detection since it serves as a launching pad for understanding more general stochastic stopping time problems involving interacting social agents; see \citeA{Kri16,KB16,Kri21}.

Classical quickest detection deals with detecting a change in an underlying state given noisy observations. We consider the following generalization: \textit{How to detect a  change in the  {strategy} of rationally-inattentive anticipatory agents when they interact  over a social network?} 
  The reader should keep in mind the following motivating example. Consider a social media accommodation system such as Airbnb \cite{EFM16}.  By observing  the reservation decisions of  customers, how can Airbnb  detect if there is a sudden change in the demand for a specific  accommodation (apartment) due to the presence of a new competitor or change in quality of the existing accommodation?  \citeA{Kri21} contains  several  results. Unlike classical quickest detection, we only have access to the actions of the agents from their  sub-game  Nash equilibrium~(\ref{eq:BNE}).
As a result, the Bayesian belief (posterior) update structure is much  more complex than  classical quickest detection. This causes remarkably counter-intuitive behavior as we will investigate below.

Quickest detection aims to
determine the 
optimal stopping time policy $\mu^*$ to minimize a cumulative cost involving false alarm and delay called the Kolmogorov--Shiryaev
criterion of disorder 
\cite{PH08,Shi63,VB13,TM10}:
\beq \inf_\mu  J_\mu(\belief_0) =   d \,\Ep\{(\tau - \tau^0)^+ | \belief_0\} +  f \, \Ep\{ I(\tau < \tau^0 ) | \belief_0 \}.
\label{eq:ksd} \eeq
Here $\tau$ is the stopping time when the detector announces a change has occurred, $\tau^0$ is the actual change time, and $\belief_0$ denotes the prior.
Waiting too long to announce a change incurs a delay penalty $d$ at each time instant after the system has changed, while declaring
a change before it happens, incurs a false alarm penalty~$f$.

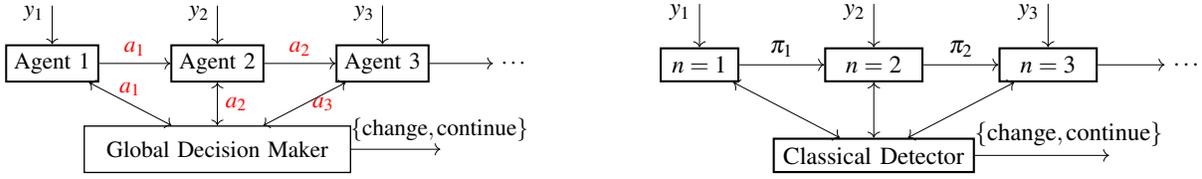
\begin{figure}[h] 

  \mbox{} \vspace{-0.3cm}
  
 \resizebox{7.5cm}{!}{       \begin{tikzpicture} [node distance =2.5cm and 2cm, auto]
                  \tikzset{every node}=[footnotesize=\small]
                  \node [blocka] (l2) {Agent 1};
                  \node [blocka,right of=l2] (l3) {Agent 2 };
                   \node [blocka,right of=l3] (l4) {Agent 3 };
                   \node [right of=l4,node distance=2cm] (cdots)[draw=none]{$\cdots$};
                   \node[above of = l2,node distance=1cm] (ar1)[draw=none]{};
                   \node[above of = l3,node distance=1cm] (ar2)[draw=none]{};
                   \node[above of = l4,node distance=1cm] (ar3)[draw=none]{};
    \node [blockd,below of=l3,node distance=1.3cm] (global) {Global Decision Maker};
    \node [right of=global,node distance=3.5cm] (globalaction)[draw=none]{};
    \node [above of=l2,left of =l2,node distance=1cm] (s2)[draw=none]{};
    \draw[->](global) -- node[above,pos=0.99]{$\{\text{{change}}, \text{{continue}}\}$}  (globalaction);
    \draw[->] (l2) --  node[above,pos=0.5]{\red{$a_1$}} (l3);
      \draw[->] (l3) --  node[above,pos=0.5]{\red{$a_2$}}  (l4); \draw[->](l4) -- (cdots);
    \draw [<->] (l2) --  node[above,pos=0.5]{\red{$a_1$}} (global);
    \draw [<->] (l3) -- node[right,pos=0.5]{\red{$a_2$}} (global);
    \draw [<->] (l4) -- node[right,pos=0.5]{\red{$a_3$}} (global);
    \draw[->](ar1) --  node[left,pos=0.2]{$y_1$}  (l2);
    \draw[->](ar2) --  node[left,pos=0.2]{$y_2$}  (l3);
     \draw[->](ar3) --  node[left,pos=0.2]{$y_3$}  (l4);
   \end{tikzpicture}} \hspace{0.9cm}
    \resizebox{8cm}{!}{       \begin{tikzpicture} [node distance =2.5cm and 2cm, auto]
                  \tikzset{every node}=[footnotesize=\small]
                  \node [blockss] (l2) {$n=1$};
                  \node [blocka,right of=l2] (l3) {$n=2$ };
                   \node [blocka,right of=l3] (l4) {$n= 3$ };
                   \node [right of=l4,node distance=2cm] (cdots)[draw=none]{$\cdots$};
                   \node[above of = l2,node distance=1cm] (ar1)[draw=none]{};
                   \node[above of = l3,node distance=1cm] (ar2)[draw=none]{};
                   \node[above of = l4,node distance=1cm] (ar3)[draw=none]{};
    \node [blockssd,below of=l3,node distance=1.3cm] (global) {Classical Detector};
    \node [right of=global,node distance=3.5cm] (globalaction)[draw=none]{};
    \node [above of=l2,left of =l2,node distance=1cm] (s2)[draw=none]{};
    \draw[->](global) -- node[above,pos=0.7]{$\{\text{{change}}, \text{{continue}}\}$}  (globalaction);
    \draw[->] (l2) --  node[above,pos=0.5]{$\belief_1$} (l3);
      \draw[->] (l3) --  node[above,pos=0.5]{$\belief_2$}  (l4); \draw[->](l4) -- (cdots);
    \draw [<->] (l2) --  (global);
    \draw [<->] (l3) --  (global);
    \draw [<->] (l4) --  (global);
    \draw[->](ar1) --  node[left,pos=0.2]{$y_1$}  (l2);
    \draw[->](ar2) --  node[left,pos=0.2]{$y_2$}  (l3);
     \draw[->](ar3) --  node[left,pos=0.2]{$y_3$}  (l4);
      \end{tikzpicture} }
    \caption{\small  Quickest Detection Problem involving multiple anticipatory agents in a line network and a global decision maker. Unlike classical quickest detection,
      the global decision maker  only observes the anticipatory actions $a_1,a_2,\ldots$.}
  \label{fig:schematic}

\end{figure}

\noindent {\bf  {Optimality of  Multi-threshold Policies in Multi-agent Quickest Detection}.}
Consider  the following   generalization of classical quickest detection.
Suppose  anticipatory agents interact
over a line network shown in Figure \ref{fig:schematic}.
The global decision maker
only observes actions $a_1,a_2,\ldots,$ from the sub-game Nash equilibria  of rationally-inattentive anticipatory  agents that influence each
other. \textit{How can the global decision maker achieve quickest detection?}  This is unlike classical quickest detection where the detector has access to  observations $y_1,y_2,\ldots$ or equivalently,  beliefs $\pi_1,\pi_2,\ldots$,
see Figure \ref{fig:schematic} (right).

\begin{figure}[h]
\mbox{} \vspace{-0.5cm}
  
\subfigure[Quickest Detection (Anticipatory)]{
\resizebox{5.5cm}{!}{ \pgfplotsset{compat = 1.3}
   \begin{tikzpicture}
\begin{axis}[width=6cm,height=3cm,ticks=none,
    ,xlabel= {\small belief $\belief$},y label style={at={(0.1,1.1)}},ylabel style={rotate=-90}
    ,ylabel={\small$\globalpolicy^*(\belief)$}
    ,axis x line = bottom,axis y line = left
    ,ymax=0.8 
    ,ymin=-1.4
    ]
    \addplot[blue,ultra thick] coordinates {(-1.1,-0.1) (-0.2,-0.1) (-0.2,-1) (0.4,-1)
      (0.4,-0.1) (0.7,-0.1) (0.7,-1) (1.1,-1)};
  \end{axis}
  \node[text width=3cm] at (0.8,1) {cont};
  \node[text width=3cm] at (0.8,0.2) {\red{stop}};
  \draw[red,<->]    (1.8,1) -- (3,1) ;
   \draw[red,<->]    (3.6,1) -- (4.5,1) ;
 \end{tikzpicture}}}
 \hspace{4cm}
\subfigure[Classical Quickest Detection]{\resizebox{5.5cm}{!}{ \pgfplotsset{compat = 1.3}
  \begin{tikzpicture}  
\begin{axis}[width=6cm,height=3cm,ticks=none, y label style={at={(0.1,1.1)}},ylabel style={rotate=-90}
    ,xlabel={\small belief $\belief$}
    ,ylabel={\small $\globalpolicy^*(\belief)$}
    ,axis x line = bottom,axis y line = left
    ,ymax=0.8 
    ,ymin=-1.4
    ]
\addplot[blue,ultra thick] coordinates {(-1.1,0) (-0.2,0) (-0.2,-1) (1.1,-1)};
\end{axis}
  \draw[red,<->]    (1.8,1) -- (4.3,1) ;
\end{tikzpicture}}}  %
%
%
%

\vspace{-0.4cm}

\caption{(a) The remarkable  multi-threshold structure of quickest detection  policy $\globalpolicy^*$ as a function of  belief (posterior)  $\belief$ in quickest detection with anticipatory agents. The stopping set is non-convex (disconnected) as shown in red.
  (b)~The optimal policy in classical quickest detection is a single threshold with convex stopping set.}
\label{fig:qdintro}
\end{figure}
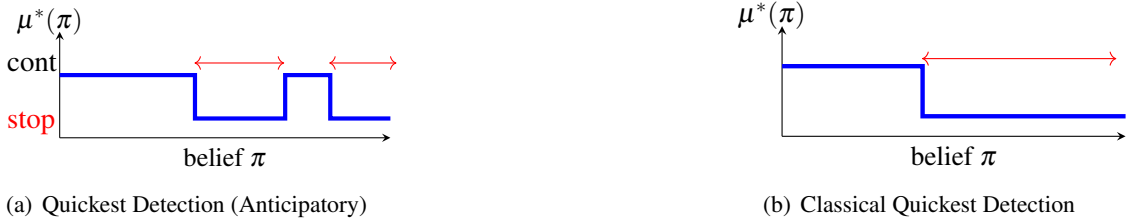

 The quickest detection policy is the solution to a stochastic dynamic programming problem in the belief state (and is an instance of partially observed stopping time problem). 
 \citeA{Kri21} shows that  the  quickest detection policy has a  {multi-threshold structure}. This is shown in Figure \ref{fig:qdintro}(a).  This is in  contrast to classical quickest detection
\cite{Shi63,PH08}, where the optimal  policy is  a  single threshold
as show in Figure \ref{fig:qdintro}(b); see also \cite{Tsi93,TV84,AT96,VV01,VV97} for  non-standard decentralized  detection results. 

Multi-threshold policies (non-convex stopping regions) are highly unusual in stochastic control.
The reason for the multi-threshold behavior in Figure \ref{fig:qdintro}(a) is discontinuity
of the Bayesian belief (posterior), i.e., 
two arbitrarily close priors can result in vastly different posteriors;
see 
\citeA{Kri21,Kri12} for a detailed analysis.

\begin{figure}[h] \centering
 \resizebox{14cm}{!}{  \begin{tikzpicture} [node distance =2.5cm and 2cm, auto]
                  \tikzset{every node}=[footnotesize=\small]
                  \node [blockff] (b1) {$\obs_n \sim \pdf(\obs| \state_n) $ observation};
                  \node [blockfg,right of=b1,node distance=4cm] (b2) {$ \private_{\dtime} = \filter(\public_{\dtime-1}, \obs_{\dtime})$ private belief update};
                  \node [blockfg,right of=b1,node distance=4cm] (b2) {$ \private_{\dtime} = \filter(\public_{\dtime-1}, \obs_{\dtime})$ private belief update};
                  \node [blockfg,right of = b2,node distance=4.3cm] (b3) {$\red{a_n} \sim$ Nash Equilibrium  anticipatory action};
                  \node [blockfg,below of = b3,node distance=1.5cm] (b4) {$ \belief_{n} = \filterg(\belief_{n-1},\red{a_{n}}) $ public belief update};
                  \node [blockfg,right of = b4,node distance=4.2cm] (b5) {Global decision $\in$ \{change, continue\}};
                  \draw [->] (b1) --  (b2);
                  \draw [->] (b2) --  (b3);
                  \draw [->] (b3) --  (b4);
                  \draw [->] (b4) --  (b5);
                  \draw [->] (b4) --  node[anchor=west] {} ++(-2,0) -| (b2);     
                \end{tikzpicture}}
                \caption{Complex  dynamics in belief (posterior) update for  Quickest Detection with anticipatory agents.}
                \label{fig:complexbelief}
              \end{figure}
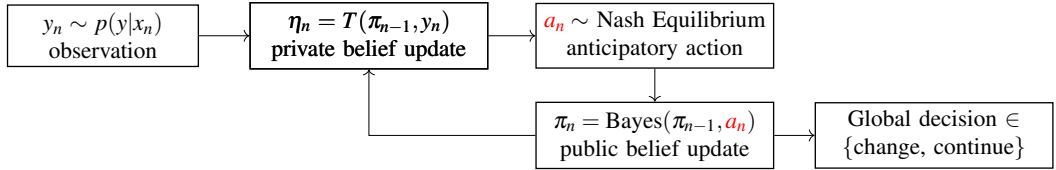

{\em Next we will study the complex  dynamics of the belief update in quickest~detection with anticipatory agents.} 
%
Note  the feedback loop in Fig.\ref{fig:complexbelief}  connects the private belief to the Nash equilibrium and public belief.
This is in comparison to the simple Bayes update $\belief_n = \filter(\belief_{n-1},y_n)$ in classical quickest detection.

              Keeping in mind the complexity of the posterior belief update in Figure \ref{fig:complexbelief}, our research tasks include: \\              (i) How to characterize the multi-threshold structure of the quickest detection policy
in Fig.\ref{fig:qdintro}(a)?  \\ (ii) How does the policy depend on the rational inattention
              scale factor $\lambda$ in (\ref{eq:ra})?   How to  generalize  the anticipatory model using the subjective belief multi-horizon formulation of \cite{BPP17}? \\
              (iii) How to generalize quickest detection  to phase distributed change times \citeA{Kri11} that are non-geometric? 



\noindent {\bf   {Bayesian Analysis. From Non-commutativity of Blackwell Dominance
    to Lehmann Precision.}}

Figure~\ref{fig:complexbelief} shows that the action $a_n$ is a noisy version of the belief $\eta_n$ (and that too with complex dynamics).
\textit{How to bound the performance of the multi-threshold quickest detector involving anticipatory agents?}
We can  use \textit{Blackwell dominance} \citeA{BK19,Kri16,Kri12},\cite{Bla53} to show that the optimal
cost achieved by the quickest detector can be lower bounded by any classical quickest detector whose observation matrix Blackwell dominates
the action likelihood in Figure~\ref{fig:complexbelief}.
   {An observation kernel $B_1$ Blackwell dominates
    $B_2$, denoted as  $B_1 \gR B_2$,  if $B_2 = B_1 Q$ for some stochastic kernel $Q$}. Intuitively $B_2$ is more noisy than~$B_1$.

 Next, suppose the anticipatory agents are organized in a \textit{hierarchical}  network (instead of the line network in Figure \ref{fig:schematic}). Then Blackwell dominance does not necessarily hold. 
To explain this, suppose level $l$ of the  network receives the underlying state distorted by a confusion matrix $Q^l$.
Then the likelihood matrix  at level $l$ is  $Q^l B$ where $B$ is an observation  likelihood matrix.
 {The key  point is that  $B\gR B Q^l$ does not imply
  that  $B \gR Q^l B$}, since the product $Q^l B$ does not commute; see  \cite{Rag11} for an interpretation in terms of Shannon capacity. {\em Put simply,
Blackwell dominance is non-commutative.}

 \citeA{Kri19} shows  for this non-commutative case, \textit{Lehmann (integral precision)  dominance} \cite{Leh88,GP10} can be used to construct 
 lower bounds to the optimal cost of quickest detection in a hierarchical network.  We can  extend  these Lehmann precision methods to quickest detection of hierarchical anticipatory agents.

\noindent {\bf  {\em  How un-informed local decision makers affect global   decision making?}}
A well known characteristic of the sequential multiagent framework is that agents herd and form information cascades
\cite{Cha04},\citeA{Kri12,KP13}. That is, agents ignore their own observations and parrot decisions of previous agents. How does herding affect the quickest detection policy? Can rational inattention (controlling the precision of the observation) delay herding? In future work is  worthwhile exploring  how  the Nash equilibrium of individual  anticipatory decision makers affects the onset of information cascades. 

\subsubsection*{Remark. Quantum Decision Theory Models}
 The decision making models we considered in this section involve classical probability.  Quantum Decision Theory (\cite{BB12a,Khr10,YS10,YS16}  and references therein) has emerged as a new paradigm which is capable of generalizing current models and accounting for certain violations of axiomatic assumptions. For example, it has been empirically shown that humans routinely violate Savage's 'sure-thing principle' \cite{KH09,ABC11}, which is equivalent to violation of the law of total probability, and that human decision making is affected by the order of presentation of information \cite{TB11,BPF11}  ("order effects"). These violations are natural motivators for treating the decision making agent's mental state as a quantum state in Hilbert Space; The mathematics of quantum probability was developed as an explanation of observed self-interfering and non-commutative behaviors of physical systems, directly analogous to the findings which Quantum Decision Theory (QDT) aims to treat. Indeed, the models of Quantum Decision Theory have been shown to reliably account for violations of the 'sure-thing Principle' and order effects \cite{BB12}. We refer to~\citeA{SKS22a} for quickest detection involving a quantum decision agent.

\section{Part 3. Interactive Sensing in  Large  Networks}

Parts  1 and 2   focussed on sequential Bayesian social learning (on a line  graph)  thereby facilitating analysis of information patterns and structural results for controlled sensing. In Part  3,  the main ideas involve statistical signal processing
 of social sensors that interact on large (possibly random)
graphs.  
The key difference compared to Parts  1 and 2 is that the methodology is primarily  non-Bayesian and the analysis is asymptotic.

Our aim is to compare how statistical signal processing methods and fundamental bounds  operate on
 two canonical network models: the  Erd\H{o}s  R\'enyi   graph~\citeB{ER60}  and the Power Law  model~\citeB{BA99}.

\subsection{Adaptive Estimation of  Degree Distribution of Evolving Random Graph}
We consider  adaptive estimation of the  degree distribution of  non-stationary random graphs. Specifically, we  consider \textit{Markov-modulated duplication-deletion random graphs}
where at each time instant, nodes can either be added to or eliminated from the graph with  probabilities that evolve according to a finite-state Markov chain;
see \citeA{HKY14}, \cite{CL06}  for a formal construction.
 Such graphs mimic social networks  \citeB{Jac10} where the interactions between nodes evolve over time according to a Markov process that undergoes infrequent jumps.
Estimating the degree distribution of a large scale graph is useful in
diffusion of information (technology,  disease) in social networks \citeB{Pin08,Pin16,Veg07,Jac10} and the existence of  ``giant components''.

Regarding adaptive degree distribution estimation:
\begin{compactenum}
\item The formulation of a two-time scale Markov modulated infinite graph with degree distribution on a Hilbert space.  
Intuitively,  since the degree distribution of a power law network dies away only polynomially, formulation on a denumerable state space (and therefore a
Hilbert space) is natural.  (There are deeper characterizations in empirical process theory in terms of the Glivenko-Cantelli class \cite{VW96} 
for weak convergence of functionals that we will not discuss.)
\item A stochastic approximation algorithm is proposed to track the evolving  stationary degree distribution of the Markov modulated
infinite random graph.
\end{compactenum}
 In practical terms, these  results  yield
  bounds on  tracking algorithms in fast changing environments.\

\subsubsection{Background: Stochastic Approximation on a Hilbert Space}

The  model of a sampled  Markov modulated  duplication deletion random graph has three ingredients: 
\begin{myenv}
 (i) {\em Stationary Degree Distribution}:  Let $\th$ denote the parameters of the duplication deletion model. 
Let $D(\th)$ denote the  stationary degree distribution of the resulting  infinite graph with support on the set of non-negative integers;
its elements are denoted by $D^{(i)}(\th),  i =0,1,2,\ldots$. 
Of course, the degree distribution is a probability mass function, i.e., $D^{(i)}(\th) \geq 0$ and $\sum_i D^{(i)}(\th) = 1$.
As mentioned above, it is convenient to imbed the denumerable vector $D(\th)$ in the Hilbert space
$\ell_2=\{z \in \reals^\infty: \sum^\infty_{i=0} \|z_i\|^2 < \infty\}$. 
\\
(ii) {\em Slow Markov chain}: Assume that the parameters of the graph $\th$ evolve according to a slow Markov chain $\{\th_k\}$, $k=0,1\ldots$.
The   transition matrix is $A^{\emc}  = \identity + \emc Q$, where $Q$ is a generator and $\emc$ is a small
positive scalar. This means that the Markov chain jumps infrequently on the time scale $k$.
Let $\{1,\dots, \statedim\}$ denote the states of this slow  Markov chain $\theta_k$.
So corresponding to each  state of this Markov  chain  is the stationary degree distribution $D(\theta)$, $\theta \in \{1,2,\ldots\statedim\}$.
\\
(iii) {\em Observation process}: At each time $k$,  nodes are sampled uniformly 
and their  degree $\observ_k$ recorded. Let $\obs_k = \mathbf{e}_{\observ_k}$ denote the observation vector where $\mathbf{e}_i $ is the $i$-th standard unit vector. Such a sampling procedure can be time correlated implying that the observations $\obs_k$ can  be a     stationary $\phi$-mixing
process \cite{EK86}. \end{myenv}

{\bf Aim}:  How can the stationary degree distribution  $D(\theta_k) \in \ell_2$ be adaptively estimated?

 \citeA{HKY14}  analyzed  the following stochastic approximation algorithm  
(where $\esa>0$ is the  step size):  
\beq \hg_{k+1} = \hg_k +\esa\left( \mathbf{e}_{\observ_k} - \hg_k\right),\quad  \text{ where } \observ_k \sim D(\th_k). \label{eq:sadegree} \eeq

Here $\hg$ is the estimated degree distribution. 
Note that the stochastic approximation algorithm (\ref{eq:sadegree}) does not
assume any knowledge of the Markov-modulated dynamics of the graph.  The
Markov chain assumption for the random graph dynamics is used only in the
convergence and tracking analysis. 

  \citeA{HKY14}  assumes that the Markov chain $\th_k$ evolves on a slower
time scale than the dynamics of the stochastic approximation algorithm, i.e., its transition matrix is $A^\rho = I + \rho Q$ where $\rho = o(\epsilon)$; for example $\rho = \epsilon^2$,
Based on the estimates $\hat{D}_k$ generated by the algorithm, define the continuous time interpolated process $\hat{D}^\epsilon(t) = \hat{D}_k $ for $t\in [k\epsilon, (k+1)\epsilon)$.
Then {\em stochastic averaging theory} \cite{BMP90,KY03,EK86}  says that   as $\epsilon \rightarrow 0$, 
the process $\hat{D}^\epsilon(t) $ 
converges weakly (weak convergence is a function space generalization of convergence in distribution \cite{EK86}; the function space here is $D([0,\infty): \ell_2)$, the space
of cadlag functions on the $l_2$ Hilbert space)  to  the ordinary differential equation (ODE);  see \citeA{HKY14} for technical details:
\beq  \frac{d\hg(t)}{dt} =    D(\th_k)  
- \hg(t) . \label{eq:classicalode} \eeq
Note that $  D(\th_k)$ is a constant in (\ref{eq:classicalode}) since it evolves on the slow time
scale $k$.
So the differential equation (\ref{eq:classicalode}) has an attractor at $  D(\th_k)$. Thus, algorithm (\ref{eq:sadegree}) converges to the true degree distribution $D(\th_k) $.

\subsubsection{Tracking a fast evolving degree distribution on Hilbert Space}
We   consider real time estimation where the Markov chain evolves at the same time scale as the  stochastic approximation algorithm: so 
the Markov process
$\theta_k$ has transition matrix $A^\rho = I + \rho Q$ where
$\rho = O(\epsilon)$  instead of the  case $\rho = o(\epsilon)$ described above. 

 Most  existing literature  analyzes stochastic approximation algorithms when the parameter  evolves according to a ``slowly time-varying''
sample path of a continuous-valued process so that
the parameter changes by small amounts over small intervals of time \cite{Mou98,GLW97,BMP90}.
 In comparison, we  cover the case where the underlying parameter evolves with discrete jumps that can be arbitrarily  large in magnitude on short intervals of time
 (same time scale as the speed of adaptation  of the stochastic approximation algorithm)
\citeA{YKI04,KTY09,YIK09,NKY17}.

\begin{thm}[Stochastic approximation convergence on Hilbert space]
\begin{compactenum} \item
 Define the tracking error $\tg_k = D(\th_k) - \hg_k$. Then, $\lim_{k \rightarrow \infty} \E\|\tg_k\|^2 = O(\varepsilon)$, where $\varepsilon$ is the step size.
\item 
 Also,  the interpolated process $\hg^\e(t) = \hat{D}_k$, $t \in [k\epsilon, (k+1)\epsilon)$ is tight in the function space $D([0,\infty): \ell_2)$.
As a consequence, as $\epsilon \rightarrow 0$,  $\hg^\e$ converges weakly to
the switched Markovian differential equation
\beq \label{ode}  \frac{d\hg(t)}{dt} = D(\th_t)  - \hg(t) , \qquad  \text{ where Markov chain } \theta_t \text{ evolves with generator }  Q. \eeq  \end{compactenum}
\end{thm}
The proof  uses  the deep ideas in the  ``martingale problem of 
 Stroock and Varadhan'' on
the Hilbert space  $\ell_2$  (see \citeB{EK86,KS85,KY03}); see our papers \citeA{YKI04,KTY09,YIK09,NKY17}
for extensive results in Euclidean space.

 The interesting property of the above theorem is that unlike   (\ref{eq:classicalode}), the limit system 
(\ref{ode}) is no longer a  deterministic 
ordinary differential equation,  but a differential equation modulated by a continuous-time Markov chain $\theta_t$. That is, the averaged system has stochastic dynamics.

Also, defining the scaled tracking error process $\nu_n = (\hg_n - D_n)/\sqrt{\epsilon}$, we will  show that the continuous time interpolated version of $\nu_n$
satisfies a functional central limit theorem similar to \citeA{YKI04}. 

{\bf \em Engineering relevance.}
  (\ref{ode}) specifies the tracking performance  when the underlying degree distribution has fast dynamics.
The covariance of the  diffusion process $\nu$ specifies the asymptotic convergence rate of the tracking algorithm \cite{KY03}. This covariance operator is on  Hilbert space $l_2$; and the variance is well~defined.


\subsection{Statistical Signal Processing for Infected Degree Distribution}
We wish to  adaptively estimate the 
{\em empirical distribution of actions} of all social sensors as they interact over a network. This empirical distribution given the degree of a node is called  the
{\em infected degree distribution}.
 The ordinary differential equation approach  will
can be used as a generative model (mean field dynamics) 
for the evolution of the infected distribution as the infection propagates over the network.
Then we will develop Bayesian signal processing algorithms to estimate the infected degree
distribution as it evolves over time. Tracking the infected degree distribution has important applications in spread of  technology, computer viruses
and strategic choices \citeB{Pin08}. 

\subsubsection{Background: Mean Field Dynamics as a Generative Model}
In Parts  1 and 2 studied  how the {\em individual actions}  of  interacting agents evolve. In comparison,
here we  estimate how the empirical distribution of actions  of {\em all} agents evolve as they interact on a network.
For simplicity, we consider the 
{\em Susceptible-Infected-Susceptible (SIS)} \citeB{Veg07,NPP15} model
which assumes each agent chooses one of two possible actions, 0 (not infected)  or 1 (infected).
In analogy to social learning, the action chosen by an agent  at each time depends probabilistically on its degree, the  action distribution of its
neighbors, and the cost it seeks to minimize. The transition probabilities specifying these actions  are denoted as $\bar{p}_{01}$ and $\bar{p}_{10}$.

The empirical distribution of the fraction of agents with degree $d$ who choose action 1  is called the
``infected degree distribution" denoted by $\rho(d)$, 
 $d\in\{1,2,\dots,\bar{D}\}$. The  following result is well known: 
\begin{thm}[Mean field dynamics \citeB{BW03,Pin08}] For degrees $d=1,2,\ldots, \bar{D}$, the infected distribution evolves as\\
\begin{minipage}{4cm} \red{(SIS Dynamics)}  \end{minipage}  \begin{minipage}{12.4cm} \begin{equation} \label{eq:minc}
\rho_{k+1}(d) = \rho_k(d) + \frac{1}{N}
\bigl[\bar{p}_{01}-\bar{p}_{10}+ w_{k+1}  \bigr]
\end{equation} \end{minipage}\\
where $N$ is the number of nodes, and $\{w_k\}$ is a martingale increment process. 
Then the mean field dynamics are given by the deterministic  difference   equation\\
\begin{minipage}{4cm} \red{(Mean Field Dynamics)}  \end{minipage}  \begin{minipage}{12.4cm}   \beq  
\brho_{k+1}(d) = \brho_k(d) + \frac{1}{N} \bigl[\bar{p}_{01}-\bar{p}_{10}\bigr]
 \label{eq:mfd}\eeq
\end{minipage}\\
where $\bar{p}_{01}$ and $\bar{p}_{10}$ are polynomial functions of the degree distribution $\brho$.  
The approximation error is
\begin{minipage}{4cm} \red{(SIS vs MFD)}  \end{minipage}  \begin{minipage}{12.4cm}  \beq \prob \bigl\{ \max_{0 \leq k \leq \finaltime} \left\| \brho_k - \rho_k \right\|_\infty \geq \epsilon\bigr\}
\leq  C_1  \exp(-C_2 \epsilon^2 N) \label{eq:azuma}\eeq\end{minipage}
 \label{thm:mfd}
\end{thm}
Theorem~\ref{thm:mfd} says that a large state-space Markov chain in (\ref{eq:minc}) for SIS can be approximated by a $\bar{D}$-dimensional  mean field difference equation and the approximation error dies exponentially in  the number of nodes $N$ (\ref{eq:azuma}).
Thus the mean field dynamics (\ref{eq:mfd})  yields a  tractable generative model \cite{PV01,Pin06,Pin08}.  

Our first task is to
  generalize the mean field dynamics (\ref{eq:mfd})  to  non-stationary Markov switched networks where $\bar{p}_{01}$ and $\bar{p}_{10}$
  are modulated by a finite state Markov chain \citeA{KNH14}.
   A similar bound
to (\ref{eq:azuma}) can  be obtained using the Azuma Hoeffding inequality for martingales.

\subsubsection{Tracking the infected distribution.  Bayesian  filtering}
Using the above mean field generative model,
{\em how to  estimate   the evolving infected degree distribution for  large networks, when the  population is sampled 
 to gather noisy information?}  This   can be posed as a  Bayesian state filtering problem where the underlying state (infected degree distribution) evolves according to the mean field dynamics (\ref{eq:mfd}); see  \citeA{KBP17}.
The key point  is that the mean field dynamics (\ref{eq:mfd})
 yield a  system whose state (infected degree distribution of network) evolves with polynomial dynamics. 
Also, by central limit theorem arguments,  the sampling (measurement)  noise of the infected degree distribution  is Gaussian. 
  Therefore,  the  filtering results in  \citeB{HB14}     for Gaussian systems with polynomial dynamics are  applicable! This  yields a  finite dimensional
  optimal  filtering algorithm  to compute the conditional mean estimate of the infected degree distribution; see \citeA{KBP17}.
  
With sophisticated  network sampling methods such as MCMC based respondent driven sampling \cite{Hec97,Hec02,GS09}, the observed infected degree distribution noise variance is an
explicit  function of the infected degree distribution.
We then have a nonlinear filtering problem without a closed form finite dimensional optimal filter.

\subsubsection{Posterior Cramer Rao bounds (PCRLB) -- Power Law vs Erd\H{o}s R\'enyi Network}   {\em How sensitive are the filtered estimates of the infected degree distribution to the underlying
network structure?} We can compute  PCRLBs  for the  mean square error filter performance \citeB{TMN98} for tracking the infected degree distribution.
 \citeA{KBP17} shows interesting behavior.
Fig.\vref{fig:pcmse}  shows the PCRLB for a power-law network with degree distribution $ \degdist(l) \propto  l^{-\gamma}$, where  $\gamma =2.7$;
and the PCRLB for an  Erd\H{o}s R\'enyi network with  degree distribution $ \degdist(l) \propto e^{-\lambda l } $, where $\lambda=2.7$. The value $\gamma=2.7$ was chosen since it is similar to the  out-degree of the World Wide Web, see \citeB{BKM00}. The displayed mean square errors in Fig.\ref{fig:pcmse} are  averaged over 100  simulations. 
The crucial point from 
Fig.~\ref{fig:pcmse}, is that  both the PCRLB and its slope are insensitive to the underlying network structure.
These suggest that for tracking the infected degree distribution, precise knowledge of the underlying network distribution is not required. 

{\bf \em Why?} The key point is that the infected link probability $\th$  (namely, the probability than a uniformly chosen link points to an infected node)  is a sufficient statistic for the  mean field dynamics model. In turns out that $\th$ is insensitive to the underlying graph structure when considering infected nodes of low degree.
This in turn implies that $\bar{p}_{01}$ and $\bar{p}_{10}$ in (\ref{eq:mfd}) are insensitive so that the PCRLBs are very similar.

 With the  results in \citeA{KBP17}, one can develop
an analytical characterization
of the sensitivity of the PCRLB with respect to network parameters; and
simulation based gradient estimation algorithms for estimating the sensitivity \citeA{KV12}. 
It is interesting to note that the variance of the maximum likelihood estimate for Erd\H{o}s R\'enyi  parameter $\lambda$ is 
substantially smaller than that for the power law parameter $\gamma$ -- the classical CR bound is almost 10 times smaller in the regions of interest.

\subsubsection{Learning Correlated Equilibria -- Differential Inclusion for Mean Field Dynamics}  {\em How can social sensors achieve coordination  in decision making?}
We  consider the following game-theoretic extension of the SIS model. Suppose social sensors in a network
 choose their action by minimizing their regret \cite{HM00}  by deploying a stochastic approximation  algorithm similar
 to  (\ref{eq:sadegree}). {\em How to characterize the global empirical distribution of the actions taken by all the agents?}
It turns out that the empirical distribution of actions converges to a convex polytope of  correlated equilibria of a repeated game, see the pioneering
works of  \cite{HM00,HM01,HMB13} -- that is simple individual behavior
by the interacting agents results in sophisticated global behavior.\footnote{Algorithms for game-theoretic learning \citeB{FL98} are broadly classified into  best response, fictitious play and regret matching.
In general it is impossible to guarantee convergence to a Nash equilibrium without imposing conditions on the structure of the utility functions
\cite{HS02,Top98}.
However, regret matching algorithms provably converge to the correlated equilibrium \cite{HM00,HM01,HMB13}.} 
Correlated equilibria  \cite{Aum87}  are a generalization of  Nash
equilibria where agents choose actions from a joint
distribution.  
Correlated equilibria are
more natural in interactive  environments than Nash equilibria since Nash equilibria assume agents act independently, which is not true
when agents interact (see   \cite{HM00,Aum87}).

 \citeA{NKY13,NKY17}  developed regret-matching stochastic approximation algorithms  that track  time-varying  correlated equilibria that evolve over time according
 to 
a  Markov process; see also 
 \citeB{BHS05,BHS06}. Using stochastic averaging theory,  the mean field dynamics  yield
  a switched Markov differential {\em inclusion} (rather than differential equation)\footnote{A 
 differential inclusion  $d\theta/dt \in F(\theta)$ specifies a family of trajectories. It  generalizes a differential equation 
 $d\theta/dt = f(\theta)$ which specifies a single trajectory. More specifically, $F$ is a Marchaud map \citeB{BHS06}, that is, the set
 $F(\theta)$ is convex and  $\sup_{y \in F(\theta)} \|y\| $ has linear growth. Differential inclusions arise in game theoretic learning since the strategies of other players
 are not known \cite{BHS06}.}; see also  consensus diffusion stochastic approximation framework \citeB{Say14b}
where  agents update  regrets  based on a linear combination of neighboring regrets; and \citeA{NKY13a,NKY13} for diffusion based
game-theoretic learning and weak convergence proofs.

\subsection{Dynamic Models for Emergence of Glass Ceiling Effect (GCE)}  
  
According to the US Department of Labor, GCE refers to \emph{``the unseen, yet unbreakable barrier that keeps minorities and women from rising to the upper rungs of the corporate ladder, regardless of their qualifications or achievements"}.
As mentioned previously,  GCE is also highly visible in social networks.
Recent works~\cite{AKL15,ADK20} study
preferential attachment models for the emergence of GCE
in   undirected networks such as Facebook.  Also, \cite{AKL15} shows that three sociological phenomena are necessary and sufficient for the emergence of GCE: {\em minority-majority partition} (i.e.~one group should be asymptotically smaller in size), {\em preferential attachment}~(i.e.~nodes should choose potential neighbors based on their current popularity)~\cite{AB02b}~ and {\em homophily}~(i.e.~nodes prefer to attach to other  nodes with similar attributes such as gender)~\cite{MSC01}. 

\noindent {{\bf Why GCE for directed networks?}}
Since  the relationship between  nodes is  asymmetric
in most social networks (Twitter, Instagram, co-citation graphs, etc),
there is strong motivation to study
GCE  in directed networks.
 \cite{SRC18,NGL16} show that in Twitter and Instagram, female users have a smaller following compared to their male counterparts. A similar empirical observation has been found in co-citation graphs  where female authors receive less attention and fewer citations compared to  male colleagues~\cite{WJK13,Iba97,SL05}. 
Our aim is to \emph{model and analyze the emergence of GCE in  dynamic directed graphs, and devise  strategies to mitigate it}. 



To explain our research plans, we first  define  GCE in a directed network. To do so, we need to quantify \textit{social influence} of  nodes in a network.
Social influence is typically quantified by  centrality measures such as degree centrality and eigenvector centrality \cite{Jac10}.
Consider a randomly
evolving network represented at time $k$ as  graph $G_k = (V_k, E_k)$.
Let 
$ \{\rednode_k, \bluenode_k\}$
denote a partition of nodes  $V_k$ into red and blue nodes.
In undirected graphs,  \cite{AKL15} uses the average degree of a group to quantify  its  social influence. However, for a directed graph, we need a metric which accounts for both the in-degree (number of followees) and the out-degree (number of followers).
We say that a network exhibits  \textit{average GCE} for blue nodes if
\begin{equation}
  \begin{array}{c} \text{{Average}} \\ \text{{GCE}} \end{array} \quad
  \boxed{\lim\sup_k\frac{\influence(\bluenode_k) }{\influence(\rednode_k) } \ll 1
  \text{ w.p.1}}
 \quad
\text{ where group influence} \quad
\influence(\bluenode_k) = \frac{\sum_{v \in{\bluenode_k}} d_o(v) }{\sum_{v \in{\bluenode_k}} d_i(v)}  \label{eq:GCE1}
\end{equation}
Here $d_o(v)$ and $d_i(v)$  denote the out and in-degrees of a node $v$.
Thus, $\influence(\bluenode_k)$ indicates the disparity between the \emph{average influence exerted by} and the \emph{average influence exerted on} a node in  group
$\bluenode_k$. For example, if $\influence(\bluenode_k) < 1$ in a citation graph, then on average,  a  member of the group $\bluenode_k$ is cited by fewer people than number of  people she cites. 

A more nuanced definition extends  GCE  to rare (tail) events - there are very few company CEOs; almost all are red, virtually none are blue. We define {\em tail GCE}
for blue as the
 existence of a glass ceiling  $\gamma$ so~that
\beq 
 \begin{array}{c} \text{{Tail}} \\ \text{{GCE}} \end{array} \quad \boxed{\lim \sup_k\frac{\proba(\influence(\text{random node} \in \bluenode_k) > \gamma)}{\proba(\influence(\text{ random node} \in\rednode_k)> \gamma)}
   \rightarrow 0} 
 \;
 \text{ where node influence } \influence(v) = \frac{d_o(v)}{d_i(v)}
 \label{eq:GCE2}
 \eeq


To fix ideas, we start with a simple  model (from our on-going work  \citeA{NAI22}) from which GCE emerges:
\begin{compactenum} \item[{\em Step 1. Birth}.] At each time instant, a new node is born with color either blue or red. The  probability of birth of a red or blue  depends on properties of the network at that time instant. 
\item[{\em Step 2. Preferential Attachment with Homophily}.]  The new node   forms {\em directed links} to existing nodes with probability  
  depending on two factors:\textit{ (i) Popularity.} The number of red and blue followers/followees the existing node has.  \textit{(ii) Homophily.} The preference of nodes to attach to nodes of its own color.
\end{compactenum}

\begin{figure}[h]
\begin{minipage}{11cm}
 	\text{\subfigure[Sample path of average GCE (\ref{eq:GCE1})]{
		\centering
                \includegraphics[width=2in] {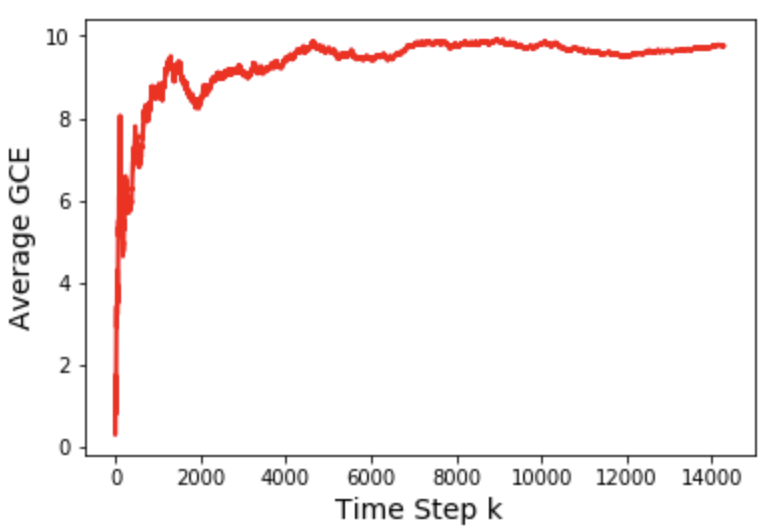}}}
            \subfigure[Snapshot of nodes indicating page-rank]{
		\centering
		\includegraphics[width=2.32in]{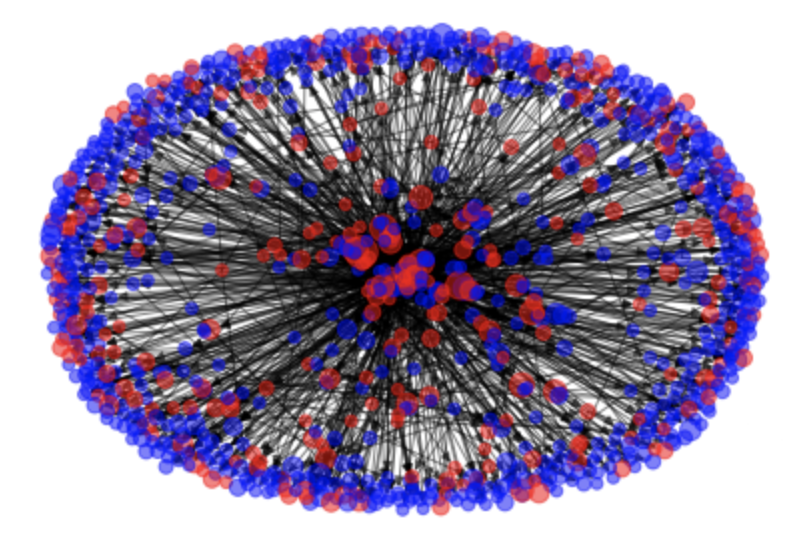}}
            \end{minipage}
            \begin{minipage}{5.4cm}
              {\footnotesize Just because blue nodes exhibit heterophily,  GCE does not necessarily emerge. One can construct examples where blue nodes
                are heterophilic and yet GCE does not emerge.
                {\em Thus GCE is   subtle and non-trivial in  directed networks
                such as Twitter and requires careful analysis.}}
            \end{minipage}

          \caption{Emergence of average glass ceiling effect  in directed network using  model in  \protect\citeA{NAI22}. 
            Fig.\ref{fig:GCE_minority_dominate}(a) shows that red nodes achieve  10 times more influence than blue nodes despite being only 20\% of the population.
In     Fig.\ref{fig:GCE_minority_dominate}(b), the size of each node  indicates its page-rank centrality. Minority  red  take on important centrality positions.}
          \label{fig:GCE_minority_dominate}
\end{figure}

\noindent
We simulated the
  above  model. In Step 1, 80\% of nodes are generated blue, 20\% are red. In Step 2, we made
blue nodes  exhibit heterophily (preference to follow red nodes), while red nodes are  unbiased in
            their  preference.
            Fig.~\ref{fig:GCE_minority_dominate} shows  emergence of GCE in the
            simulated  network:
Eventually  $90\%$ of social influence belongs to  $20\%$ of the population (red). The minority (red)
            prevents the majority  (blue) from achieving high influence.
            Another useful aspect of the model is that 
preferential attachment (Step 2)  ensures  the simulated network exhibits a power law degree distribution which mimics directed social networks such as Twitter.


\paragraph{Summary}  Part  3 discussed  social sensing on large (random) graphs. The key ideas involve adaptively estimating
degree distributions (via stochastic approximation algorithms),   tracking infected degree distributions using mean field dynamics as a generative model, and a comparison of how the graph structure affects estimates.  We also discussed how the glass ceiling effect emerges in social networks.

There are several other interesting sociological effects that can be modeled via dynamic  models in social networks. \citeA{LNK21} shows how using Markov random bridges (which are one dimensional Markov random fields), one can model echo chambers in social networks. \citeA{LNK22} also shows how segregation in social networks can be mitigated by providing incentives to agents (the formulation involves an edge formation game).

 \section{Part 4.  Polling Social Networks -- The Friendship Paradox} \label{sec:theme4}

 Our final theme in this paper 
 studies statistical estimation algorithms for polling nodes in a social network.  {We view polling  as  a
generalization of information fusion (Theme 1) with the flexibility of {\em who} to poll and {\em what} question to ask.} In large social networks, only a fraction of nodes can be polled to determine their decisions. \textit{Which nodes should be  polled to achieve a statistically accurate estimate} of phenomena such as the glass ceiling effect?
A related question is: {\em What question to ask the polled nodes?} Nodes often have \textit{social desirability bias}, i.e., they are embarrassed/reluctant to reveal their true voting intention; this can result in inaccurate poll estimates.
\textit{How to  provide incentives to  nodes so that they reveal their true~opinion?}

\noindent {\large{\bf Background.}}
Consider a  social network represented by a graph $G = (V, E)$  where each node $v\in V$ has a binary label $f(v) \in \{0, 1\}$, representing for example the intention to vote for a certain political party, or infected with a disease.
The aim of  {\em randomized polling} of a social network (with  possibly unknown structure) is
to estimate the fraction of nodes  with label 1 by polling only a subset of nodes, i.e., \begin{equation}
\label{eq:true_value}
\truevalue  = \frac{ \vert\{v\in V: f(v) = 1\}\vert  }{\vert V\vert} ,
\end{equation}

Three widely used polling strategies are \cite{RW10,DKS12}: 
\begin{compactitem} \item[{\em Intent Polling (IP)}.]  Each uniformly  sampled node is asked: {\em Who will you vote for?}
The average 
$
\hat{f} = {\sum_{u\in S}f(u)}/{\vert S \vert}$,
is used as the estimate  of the fraction $\truevalue$.  
The sample size to achieve error  $\epsilon$ is $O(\frac{1}{\epsilon^2})$.
\item[{\em Expectation Polling (EP)}.] Each uniformly sampled node is asked: {\em Who do you think will win?}
Intuitively,  EP   is more accurate\footnote{\cite{RW10} analyzes US presidential electoral college results from 1952-2008
  and shows that expectation polling was substantially more accurate than intent polling.
  The dataset from  American National
	Election Studies comprised of voter responses to two questions:
	{\em Intent Polling}: Who will you vote for in the election?
	{\em Expectation Polling}: Who do you think will be elected President?} than IP  since
each node  considers its own intent together with the intents of its friends. But this is not always true \cite{DKS12}; certain network
      structures can  increase the bias of EP.

    \item[{\em Social Sampling~(SS)}.] To reduce  the bias and variance of EP, \cite{DKS12} proposes an extension of EP  called social sampling.  The response of each sampled node is weighted by  the labels and degrees
      of its neighbors. 
\end{compactitem}
These  polling methods have  limitations. First,
EP and SS require the
pollster to have significant information about the network structure 
to reduce bias and variance.
Also, IP requires  large number of sampled nodes to achieve a desired accuracy.
Second, IP and SS do not take the privacy of the nodes into account. Attributes such as voting preferences are privacy sensitive and may lead to  people  falsely responding to the polls.

We   aim  to \textit{construct and analyze  polling algorithms based on  the friendship paradox  and strategic  querying to elicit truthful responses}. 
In future  work we will investigate controlled correlated polling strategies that optimize a multi-horizon objective. These correlated polling strategies are similar to 
 respondent driven sampling \cite{Hec97,Hec02,GS09} which is used by the US Center for Disease Control to  sample from  marginalized populations.

\subsection*{Friendship Paradox based Polling in Social Networks}
The friendship paradox was discovered by Feld \cite{Fel91} and informally states ``\textit{on average, your friends have more friends than you do}''. A nicely packaged formulation  in terms of stochastic dominance is
\cite{CR16}:
\begin{theorem} { (Friendship Paradox \cite{CR16})}
	\label{th:friendship_paradox_Feld} {\rm
	Let ${G = (V,E)}$ be an undirected graph.
	\begin{compactenum}
		\item        Let random variable $X$ denote a uniformly  chosen node from $V$.
		\item        Let random variable $Y$ denote  a uniformly chosen node from a uniformly chosen edge $e\in E$.
		\item  Let  $Z$ denote a uniformly chosen neighbor (friend) of a uniformly chosen node from $V$.
	\end{compactenum}
	Then, with $d(X)$ denoting the degree of node $X$ and  $\gs$ denoting first order stochastic  dominance \cite{MS02}
	$$  d(Y) \gs d(X)  , \quad \text{ and } d(Z) \gs    d(X)  $$
	Therefore,   the expected degree of $Z$ and expected degree of $Y$ are  larger than
	the expected degree of $X$.\qed}
\end{theorem}

\citeA{ANA20} studies  friendship paradox based polling 
in \textit{directed} graphs. Specifically, in \citeA{ANA20,NK19,NK21} we have  developed the following \textit{Network Expectation Polling (NEP)} algorithm:
\begin{compactenum} \item[\textit{Step 1 (Sampling)}.]
Choose the node $Y$ or $Z$ in Theorem \ref{th:friendship_paradox_Feld}.  Note that $Y$ and $Z$ are \textit{non-uniform} samplers of nodes
  (unlike $X$). Also $Y$ and $Z$ can be sampled without knowing the network structure (see below).

  \item[\textit{Step 2 (Querying)}.] Ask the sampled node: What fraction of your friends will vote for candidate $A$? 
  \end{compactenum}
 How to implement Step 1? 
  Sampling $Y$ in Step 1 is equivalent to sampling from the stationary distribution of a random walk on the graph \cite{Dur19}; the convergence rate depends on the second largest eigenvalue of the graph Laplacian. So the implementation of Step 1 for $Y$ is simple: just  walk through the network and compute the average of responses.
  Sampling $Z$  involves choosing a random friend of a randomly sampled node. 
  
\begin{figure} \centering
   \includegraphics[scale=1]{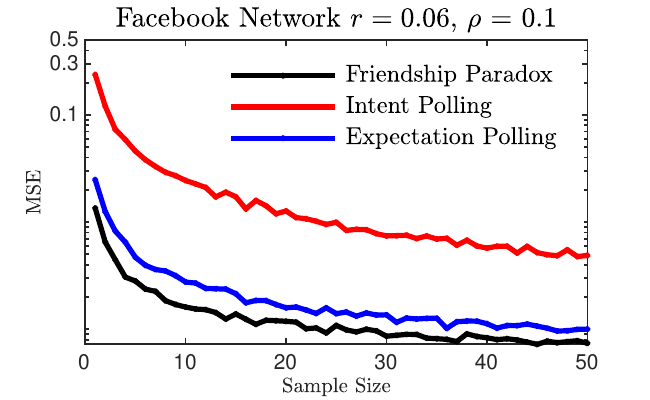}
   \caption{\small Friendship paradox based polling vs Intent and Expectation Polling (from~\protect\citeA{ANA20}).  $r$ and $\rho$ denote the assortative coefficient and   degree label correlation. The Facebook dataset is from~\cite{LM12}.} \label{fig:fpintro}
 \end{figure}

  \citeA{ANA20} proves that NEP  outperforms 
  intent and expectation polling
  in terms of mean squared error,
 recall Fig.\ref{fig:fpintro}. 
 Also 
 the sampled node  responds
 with the average of its friends' opinions (rather than its own opinion);
 {so nodes  can truthfully reveal politically incorrect opinions without personal embarrassment.}

In~\citeA{NKL19} the friendship paradox poling is used in estimating the opinions of nodes when information diffuses in social network with monophilic contagion.

\section{Closing Comments}
\label{sec:closing-comments}

This paper has surveyed four important aspects of social learning and social networks, namely, classical social learning, social learning with anticipatory agents, dynamics of social networks, and polling social networks.

One area we have not discussed is inference of optimal behavior of agents in a social network. This is driven both by technological advances such as the requirement for adaptively caching information in wireless networks \citeA{HNK15}, and also for modeling the user engagement in social multimedia networks \citeA{NHK16,HAK17,HKP20}. These  involve inverse reinforcement learning
and revealed preference.

\bibliographystyle{abbrv}
\bibliography{$HOME/texstuff/styles/bib/vkm.bib}


\end{document}